\documentclass[preprint,showpacs,preprintnumbers,amsmath,amssymb]{revtex4}
\def\pslash{\rlap{\hspace{0.02cm}/}{p}}

\def\qslash{\rlap{\hspace{0.01cm}/}{q}}
\def\kslash{\rlap{\hspace{0.02cm}/}{k}}

\def\Dslash{\rlap{\hspace{0.07cm}/}{D}}

\usepackage{bm}
\usepackage{graphicx}
\usepackage{dcolumn}

\input epsf.tex

\begin{document}

\title{
Improved $\alpha^4$ Term of the Muon Anomalous Magnetic Moment}

\author{Toichiro Kinoshita }
\email{tk@hepth.cornell.edu}
\affiliation{Laboratory for Elementary Particle Physics\\ Cornell University, 
         Ithaca, New York, 14853  }

\author{Makiko Nio}
\email{nio@riken.jp}
\affiliation{Theoretical Physics Laboratory,
RIKEN, Wako, Saitama, Japan 351-0198 }

\date{\today}

\begin{abstract}
We have completed the evaluation of all mass-dependent
$\alpha^4$ QED contributions to the muon $g-2$, or $a_\mu$,
in two or more different formulations.
Their numerical values have been greatly improved by 
an extensive computer calculation.
The new value of the dominant $\alpha^4$ term  
$A_2^{(8)} (m_\mu / m_e ) $ is 132.6823 (72),
which supersedes the old value 127.50 (41).
The new value of the three-mass term
$A_3^{(8)} (m_\mu / m_e , m_\mu / m_\tau ) $ is 0.0376 (1).
The term $A_2^{(8)} (m_\mu / m_\tau ) $ 
is crudely estimated to be about 0.005
and may be ignored for now.
The total QED contribution to $a_\mu$ is
$116~584~719.58~(0.02)(1.15)(0.85) \times 10^{-11}$,
where 0.02 and 1.15 are uncertainties in the $\alpha^4$ and $\alpha^5$
terms and 0.85 is from the uncertainty in $\alpha$
measured by atom interferometry.
This raises the Standard Model prediction by $13.9 \times 10^{-11}$,
or about 1/5 of the measurement uncertainty of $a_\mu$.
It is within the noise of current uncertainty ($\sim 100 \times 10^{-11}$)
in the estimated hadronic contributions to $a_\mu$.
\end{abstract}

\pacs{ PACS numbers: 13.40.Em, 14.60.Ef, 12.39.Fe, 12.40.Vv }

\maketitle

\section{Introduction and summary}
\label{sec:intro}

The latest measured value of the anomalous magnetic moment
of negative muon is
\cite{bennett1}
\begin{equation}
a_{\mu^-}({\rm exp}) 
  = 11~659~214~(8)~(3) \times 10^{-10} ~~~~~~~~(0.7 {\rm ~ppm}),
\label{amuexp00}
\end{equation}
where $a_\mu \equiv \frac{1}{2} (g_\mu - 2)$ and
the numerals 8 and 3 in parentheses represent
the statistical and systematic 
uncertainties in the last digits of the measured value.
$1~{\rm ppm} = 10^{-6}$.
The world average value $a_\mu (\rm{exp})$ obtained
from this and earlier measurements \cite{bennett2,brown1,brown2,CERN} is 
\begin{equation}
 a_{\mu}({\rm exp})
  = 11~659~208~(6) \times 10^{-10} ~~~~~~~~(0.5 {\rm ~ppm}).
\label{amuexpall}
\end{equation}
This result provides the most stringent test of the Standard Model.

Unfortunately, such a test must wait for further improvement in the
uncertainty of the hadronic corrections to $a_\mu$ 
\cite{davier2,ghozzi,hagiwara,ezhela,czarnecki1,davier1,narison1,troconiz1,knecht2,ramsey,melnikov}.
The lowest-order hadronic vacuum-polarization 
effect has thus far been determined from two sources,
(i) $e^+ e^-$ annihilation cross section, and (ii) hadronic
$\tau$ decays.  
Several recent evaluations are listed in Table \ref{tablehad1}.
Their differences (except for the one obtained from
the $\tau$ decay data) are due to different
interpretations and treatments of basically identical data.
However, they all agree that
the measurement of the $e^+ e^-$ annihilation cross section,
in particular in the region below $\rho-\omega$ resonances,
must be improved substantially in order to reduce
the experimental uncertainty significantly. 
Such efforts are underway at several laboratories.
Particularly interesting and promising is new radiative-return measurements
\cite{aloisio}.
On the other hand, it is not clear at present whether 
the value from the $\tau$-decay data can be improved
much further because of the difficulty in evaluating more precisely
the effect of isospin breaking \cite{davier2,ghozzi}.

A new theoretical development is an attempt to calculate the
hadronic vacuum-polarization effect on muon $g-2$
in lattice QCD \cite{blum}.

The NLO hadronic contribution has been evaluated 
by two groups \cite{krause,hagiwara}:
\begin{eqnarray}
a_{\mu}(had. NLO) &=& -10.1~ (0.6) \times 10^{-10},  \nonumber    \\
a_{\mu}(had. NLO) &=& -9.8~(0.1)_{exp}~(0.0)_{rad} \times 10^{-10} .
\label{hadnlo}
\end{eqnarray}
The contribution from radiative corrections is identical
in two papers. The small difference comes from the diagram 
in which two hadronic
vacuum-polarizations are inserted in the second-order vertex diagram.

The contribution of hadronic light-by-light scattering 
to $a_\mu$ is more difficult to obtain a reliable
value because it cannot
utilize any experimental information and must rely solely on theory.
After correction of a sign error, it seemed to
have settled down to around 
\cite{czarnecki1,davier1,narison1,troconiz1,knecht2,ramsey}
\begin{equation}
 a_{\mu}({\rm had.l-l}) \sim 80~(40) \times 10^{-11}.
\label{amuhadl-l}
\end{equation}
More recently, however, 
a considerably different value was reported \cite{melnikov}:
\begin{equation}
 a_{\mu}({\rm had.l-l}) \sim 136~(25) \times 10^{-11},
\label{newamuhadl-l}
\end{equation}
which moves the prediction of the Standard Model closer to the experiment.
This was obtained by imposing the short-distance QCD constraints on the
$\pi^0\gamma^*\gamma$ amplitude, which was overlooked in previous
analyses.  Further confirmation of this result by a first principle calculation
in lattice QCD would be highly desirable.

\begin{table}
\renewcommand{\arraystretch}{1.0}
\begin{center}
\caption{Recent evaluations of lowest-order hadronic vacuum-polarization contribution to the muon $g-2$.
Some errors are separated according to their sources: 
measurement errors and radiative corrections.
\cite{ezhela} mentions a procedural error separately.
\\
\label{tablehad1}
} 
\begin{tabular}{llr}
\hline
\hline
~~~~~process~~& ~~$a_\mu (had.LO) \times 10^{10}$~~~~&~Reference \\ [.1cm]  
 \hline
$e^+ e^-$ annihilation~~~~&~$696.3~(6.2)_{exp} (3.6)_{rad}$~&~~\cite{davier2}~~\\ [.1cm]
$e^+ e^-$ annihilation~~~~&~$694.8~(8.6)$~&~~\cite{ghozzi}~~\\ [.1cm]
$e^+ e^-$ annihilation~~~~&~$692.4~(5.9)_{exp} (2.4)_{rad}$~&~~\cite{hagiwara}~~\\ [.1cm]
$e^+ e^-$ annihilation~~~~&~$699.6~(8.5)_{exp} (1.9)_{rad}~(2.0)_{proc}$&~~\cite{ezhela}~~\\ [.1cm]
\\
$\tau$ decay&~$711.0~(5.0)_{exp} (0.8)_{rad}~(2.8)_{SU(2)}$&~~\cite{davier2}~~\\ [.1cm]
\\
\hline
\hline
\end{tabular}
\end{center}
\end{table}
\renewcommand{\arraystretch}{1}

The weak interaction effect is known to two-loop order.
The latest values are \cite{knecht3,czarnecki2}
\begin{eqnarray}
a_{\mu}(weak) &=& 152~(1) \times 10^{-11} \nonumber    \\
a_{\mu}(weak) &=& 154~(1)~(2) \times 10^{-11} ,
\label{newweak}
\end{eqnarray}
where (1) and (2) in the second line
 are the remaining theoretical uncertainty and
 Higgs mass uncertainty, respectively.
Although the numerical difference between these values
is insignificant for comparison with experiment, 
their approach to the fermionic triangle diagram
seems to be different.
We hope it is resolved before long.

The QED contribution $ a_\mu ({\rm QED})$,
even though it is the predominant term of $a_\mu$,
has received little attention thus far because of its small error bars.
The theoretical uncertainty comes predominantly from the
$\alpha^4$ term whose contribution to $a_\mu$ is about 3.3 ppm.
The best value of $a_\mu$(QED) 
reported previously  (Eq. (11) of \cite{hugheskinoshita}) was
\begin{eqnarray}
 a_\mu ({\rm QED})_{old}
  &=& 116\ 584\ 705.7\ (1.25) (1.15) (0.5) \times 10^{-11}  \nonumber  \\
  &=& 116\ 584\ 705.7\ (1.8) \times 10^{-11},
 \label{amuQED}
\end{eqnarray}
where 1.25 and 1.15 come from the uncertainties in 
the calculated $\alpha^4$ and  estimated $\alpha^5$ terms, respectively,
and  0.5 is from the uncertainty in the fine structure
constant $\alpha$ given in Eq. (17) of \cite{hugheskinoshita} 
obtained from the measurement and theory of $a_e$.

While updating $a_\mu (QED)$, however, we
 discovered that the previous evaluation of 
the $\alpha^4$ term suffered from
an error in a group of 18 Feynman diagrams \cite{kn1}.
This affects both Eq. (11) and Eq. (17) of Ref. \cite{hugheskinoshita}
so that (\ref{amuQED}) had to be revised.
This discovery prompted us to reexamine all other $\alpha^4$ terms
contributing to $a_\mu (QED)$.

The purpose of this paper is to report the result of this reexamination.
We give a full account of

\noindent
(1)  new evaluation of mass-dependent $\alpha^4$ term
of $a_\mu$ in an alternate formulation, 

\noindent
(2) vastly improved numerical precision 
by an extensive numerical evaluation
of 469 eighth-order Feynman diagrams,  and

\noindent
(3)  new evaluation of the $\alpha^4$ term that depends
on three masses $m_e, m_\mu , m_\tau$
($0.1094~(3) \times 10^{-11}$), which replaces
the old value ($0.23 \times 10^{-11}$) quoted in \cite{km}. 

If one uses the latest value of $\alpha$ obtained from the 
atom interferometry measurement
\cite{wicht}:
\begin{equation}
 \alpha^{-1} (a.i.) = 137.036~000~3~(10)~~~~[7.4~ppb],
 \label{alph_Cs}
\end{equation}
the new estimate of the QED contribution becomes 
\begin{equation}
a_{\mu}(QED) = 116~584~719.58~(0.02)(1.15)(0.85) \times 10^{-11} ,
\label{newQEDvalue}
\end{equation}
where 0.02 replaces the previous uncertainty 1.25 of 
the $\alpha^4$ term in (\ref{amuQED}),
an improvement of factor 60.
The error 0.85 comes from the uncertainty in the fine structure
constant $\alpha$(a.i.) given in (\ref{alph_Cs}).
Note that this error is larger than
the corresponding error in (\ref{amuQED}) because we used $\alpha$(a.i.)
of (\ref{alph_Cs}) instead of the incorrect 
$\alpha (a_e )$ used in (\ref{amuQED}).
The new value (\ref{newQEDvalue}) is larger than (\ref{amuQED}) 
by $13.7 \times 10^{-11}$.
Report on the improvement of $a_e$ and $\alpha (a_e)$ 
is being prepared \cite{kn3}.

As is seen from (\ref{newQEDvalue}), 
the largest source of QED error is now the $\alpha^5$ term,
which was previously estimated to be $6.29 (1.15) \times 10^{-11}$
\cite{km,karsh}.
Although this is accurate enough for comparison with the
current experimental data, a more precise value will become necessary
in the future.
It is being improved at present  and will be reported shortly \cite{knX}.

Let us now present an outline of our approach to $a_\mu ({\rm QED})$
and a summary of results before going into details.
The contribution of QED diagrams to $a_\mu$
can be written in the general form
\begin{equation}
a_{\mu}(QED)~=~A_1~+~A_2 ( m_{\mu} /m_e )~+~A_2 ( m_{\mu} / m_{\tau} )
~+~A_3 ( m_{\mu} / m_e , m_{\mu} / m_{\tau} ) ,    
\label{muanomaly}
\end{equation}
where $m_e,~ m_{\mu}$, and $m_{\tau}$ are the masses of the electron, muon, and tau, respectively.  Throughout 
this article we shall use the values $m_e$ = 0.510 998 902(21) MeV/$c^2$, 
$m_{\mu}$ = 105.658 3568(52) MeV/$c^2$, 
and $m_{\tau}$ = 1 777.05(29) MeV/$c^2$, respectively \cite{mohr}.

The renormalizability of QED guarantees that 
$A_1$, $A_2$, and $A_3$ can be expanded in power series 
in $\alpha / \pi $ with finite calculable coefficients:
\begin{equation}
A_i~=~A_i^{(2)} \left( \frac{\alpha}{\pi} \right)~+~A_i^{(4)} \left( \frac{\alpha}{\pi} \right )^2 ~+~A_i^{(6)} \left( \frac{\alpha }{\pi } \right)^3 ~+~.~.~.~,~~~~i~=~1,~2,~3.     \label{power}
\end{equation}
$A_1^{(n)}$ is known up to $n=4$ 
from the study of the electron anomaly $a_e$
\cite{hugheskinoshita,kn3}.  
$A_1^{(2)}$, 
$A_1^{(4)}$, 
and $A_1^{(6)}$ have been evaluated precisely by both 
numerical and analytic means. 
$A_1^{(8)}$ is currently being improved 
by an extensive computer work \cite{kn3}.
For the purpose of evaluating $a_\mu({\rm QED})$, 
however, we may use $A_1$
obtained from the measured value of
the electron anomaly $a_e$ \cite{vandyck}
subtracting small contributions 
due to muon, hadron, and weak interactions
\cite{alpha}.

It is easy to see that $A_2^{(2)} = A_3^{(2)} = A_3^{(4)}=0$: they have no corresponding Feynman diagram.  
$A_2^{(4)}(m_\mu /m_e)$, $A_2^{(6)}(m_\mu /m_e)$, and $A_3^{(6)}(m_\mu /m_e, m_\mu /m_\tau)$ have been evaluated 
by numerical integration, asymptotic expansion in $m_\mu /m_e$,
power series expansion in $m_e /m_\mu$, and/or analytic integration.
They are \cite{kino1,laporta1,laporta2,czarnecki}
\begin{eqnarray}   
A_2^{(4)} (m_\mu /m_e ) &=& 1.094~258~282~8~(98) ,  \nonumber  \\
A_2^{(4)} (m_\mu /m_\tau ) &=& 7.8059~(25) \times 10^{-5} ,  \nonumber  \\
A_2^{(6)} (m_\mu /m_e ) &=& 22.868~379~36~(23) ,  \nonumber  \\
A_2^{(6)} (m_\mu /m_\tau ) &=& 36.054~(21) \times 10^{-5} ,  \nonumber  \\
A_3^{(6)} (m_\mu /m_e, m_\mu /m_\tau ) &=& 52.763~(17) \times 10^{-5} ,  
\label{a2(4,6)}
\end{eqnarray}   
where the errors are due to measurement uncertainty of
$m_\mu$ and $m_\tau$ only.
The most striking feature of the $\alpha^3$ term is 
the large size of $A_2^{(6)} (m_\mu /m_e )$.
It comes predominantly from diagrams involving 
a light-by-light scattering subdiagram,
as was first discovered in \cite{aldins}
and improved by numerical calculation
\cite{km}.
Since $A_2^{(6)} (m_\mu /m_e )$ is now known analytically \cite{laporta2},
its uncertainty depends only on the uncertainty 
in the measurement of $m_e /m_\mu$ 
and is totally negligible.

The term $A_2^{(8)} (m_\mu /m_e )$ has been known 
by numerical integration only.
A crude evaluation of contributing integrals 
made more than 10 years ago \cite{km},
which was no more than an order of magnitude estimate,
showed that $A_2^{(8)} (m_\mu /m_e )$ contributes only about 3 ppm to $a_\mu$.
Thus it seemed that it
was good enough for comparison with the experiment.
Now that a program error was found in a part of evaluation of
$A_2^{(8)} (m_\mu /m_e )$ \cite{kn1} and since
the measurement of $a_\mu$ is becoming more precise,
however, it is important to re-examine 
these calculations and eliminate algebraic 
error, if any, completely and reduce the computational 
uncertainty as much as possible.

Within the Feynman gauge two approaches had been
developed for numerical integration of Feynman diagrams
contributing to the anomalous magnetic moment \cite{ck1}.
An obvious and straightforward one is to evaluate 
each vertex individually and add them up.
(This approach will be called {\it Version B} following \cite{kn1}.)
Another one starts by combining several vertices into
one with the help of the Ward-Takahashi identity
\begin{equation}
q_\mu \Lambda^{\mu} (p,q) = - \Sigma (p+\frac{q}{2}) + \Sigma (p-\frac{q}{2}),
\label{wtid}
\end{equation}
where $\Lambda^{\mu} (p,q)$
is the sum of vertices obtained by inserting the external
magnetic field in fermion lines of a self-energy diagram
$\Sigma (p)$.  $p \pm q/2$ is outgoing (incoming) muon momentum.
Differentiating both sides of (\ref{wtid})
with respect to $q_\nu$ one obtains
\vspace{2mm}
\begin{equation}
\Lambda^{\nu} (p,q) \simeq - q^{\mu} \left [ \frac{\partial \Lambda_\mu (p,q)}{\partial q_\nu} \right ]_{q=0} - \frac{\partial \Sigma (p)}{\partial p_\nu}.
\label{wtid2}
\end{equation}
Obviously one may start from either the LHS or RHS of this equation
to evaluate the anomalous magnetic moment.
The approach based on the RHS and LHS
will be called {\it Version A} and {\it Version B}, respectively.
The former required some additional algebraic work
but produced fewer integrals 
and ensured significant economy of computing time.

Evaluation of the $\alpha^3$ term was carried out 
in both {\it Version A} and {\it Version B} \cite{ck1}.
But, for the $\alpha^4$ term, in particular for 126 diagrams containing 
a light-by-light scattering subdiagram, the {\it Version B} codes 
were so large that we chose initially to work only with {\it Version A}. 
For the reason discussed already we have now reevaluated
them also in {\it Version B} \cite{kn1}.
We have now extended this effort to the remaining 108 diagrams and 
obtained their codes in {\it Version B}.
Numerical evaluation shows that they are in good agreement
with those of {\it Version A}.
As a consequence {\it all} $\alpha^4$ diagrams
contributing to $A_2^{(8)} (m_\mu /m_e )$
have been confirmed by two or more independent formulations.
We are confident that all codes are now free from any algebraic error.

The remaining problem concerns the reliability
of numerical integration.  As a matter of fact, 
values of some integrals 
were called into question
shortly after the old result was published
\cite{broadhurst}. 
It turned out that this was caused mainly by
the relatively poor statistical sampling
of the integrand  resulting from
 shortage of computing power then available \cite{kino2}.
The problem was made worse by the presence of
severe non-statistical errors that
originate from round-off errors inherent in {\it all} computer calculation.
This will be called {\it digit-deficiency errors}.  
Various techniques had to be introduced to alleviate
this problem \cite{kn2}.
See Appendix \ref{ddproblem} for details.

Now that the validity of codes is established 
we are justified to evaluate all integrals contributing to
the $\alpha^4$ term in either {\it Version A} or {\it Version B},
using vastly increased number of sampling points,
made possible by the new generations of computers,
and, at the same time, reducing {\it digit-deficiency} errors
to a manageable level by various means. 
(See Appendix \ref{ddproblem}.)

All integrals have been evaluated 
with successively increasing statistics
over the period of more than 10 years.
Some  preliminary results
were reported from time to time \cite{earlyresults}.
Only the latest and most accurate results are listed in Tables   
\ref{table1} --- \ref{tablexx}.
Although earlier results are not shown explicitly, 
they have played
crucial roles in checking the reliability of numerical
integration at every stages of calculation.

The majority of integrals in the {\it Version A} calculation
were found to be consistent with the results in {\it Version B}.
But some of them were found to differ considerably
because of poor statistical samplings and the $d$-$d$ problem.
Thus some {\it Version A} integrals have been reevaluated to reduce
the $d$-$d$ problem.
Evaluations of both versions are combined in quadrature,
whenever appropriate,
to improve the statistics.

The latest value of 
$ A_2^{(8)} ( m_{\mu} / m_e ) $ is 
\begin{equation}
A_2^{(8)} ( m_{\mu} / m_e ) ~=~132.682~3~(72),     
\label{newa28mue}
\end{equation}
which is larger by 5.2 than the old value \cite{hugheskinoshita}
\begin{equation}
A_2^{(8)} ( m_{\mu} / m_e ) ~=~127.50~(41).     
\label{olda28mue}
\end{equation}
The difference between (\ref{newa28mue}) and (\ref{olda28mue})
is partly accounted for by the correction of program error described
 in \cite{kn1} but
is mostly due to the fact that (\ref{olda28mue}) suffered from poor statistics
and the {\it digit-deficiency} problem.

There is also a small contribution to $a_\mu$ from the three-mass term
$A_3^{(8)} (m_\mu /m_e, m_\mu /m_\tau )$ which arises from 102 diagrams 
containing two or three closed loops of {\it v-p} and/or {\it l-l} type.
Results of numerical evaluation are given in 
(\ref{a3I8}), (\ref{a3II8}), and (\ref{a3IV8}).  From these results we obtain 
\begin{equation}
A_3^{(8)} ( m_{\mu} / m_e , m_\mu / m_\tau ) ~=~0.037~594~(83).     
\label{muetau}
\end{equation}
This is smaller than the value 0.079 (3) quoted in \cite{km}.
which corresponds to (\ref{a3IV8})
obtained only from the diagrams containing {\it l-l} loop,
which were thought to be dominant.
The new result (\ref{muetau}) shows that this assumption was not
fully justified.
Another term of order $\alpha^4$ is $A_2^{(8)} (m_\mu / m_\tau )$
which is calculable from 469 Feynman diagrams.
However, its contribution to $a_\mu$ is of order
$(m_\mu / m_\tau )^2 \ln (m_\tau / m_\mu ) A_2^{(8)} (1) \sim 0.005$ so that it may be safely 
ignored for now.

Collecting all results of orders $\alpha^4$ and $\alpha^5$ 
\cite{karsh} we find $a_\mu ({\rm QED})$ given in (\ref{newQEDvalue}).
In conclusion we have found that the improvement of the $\alpha^4$ term 
does not significantly 
affect the comparison of theory and experiment of $a_\mu$.
The net effect of our calculation
is to enhance the QED prediction (\ref{amuQED}) by $13.6 \times 10^{-11}$ and
eliminate an important source of theoretical uncertainty.
As far as QED is concerned, the $\alpha^5$ term
is now the most important source of uncertainty in $a_\mu$.
This is being improved \cite{knX}.
The overall theoretical uncertainty of 
the Standard Model remains dominated
by that of the hadronic vacuum-polarization effect.

\section{Classification of  diagrams contributing to $A_2^{(8)} (m_\mu / m_e )$}
\label{sec:qedcontribution}

There are altogether 469 Feynman diagrams contributing
to $ A_2^{(8)} ( m_{\mu} / m_e ) $.
Feynman integrals for these eighth-order vertex diagrams
consist of twelve propagators integrated over 4 four-dimensional
loop momenta.
These diagrams have subdiagrams of
vacuum-polarization ({\it v-p}) type and/or  
light-by-light scattering ({\it l-l}) type.
The {\it v-p} subdiagrams found in $A_2^{(8)} (m_{\mu} / m_e )$ are as follows:

\noindent
$\Pi_2$, which consists of one closed lepton loop of second-order.

\noindent
$\Pi_4$, which consists of three proper
closed lepton loops of fourth-order.

\noindent
$\Pi_{4(2)}$, which consists of three lepton loops 
of type $\Pi_4$ whose internal photon line has a $\Pi_2$ insertion.

\noindent
$\Pi_6$, which consists of 15 proper
closed lepton loops of sixth-order.

The {\it l-l} diagrams we need are:

\noindent
$\Lambda_4$, which consists of six proper
closed lepton loops of fourth-order, with four photon
lines attached to them.

\noindent
$\Lambda_4^{(2)}$, which consists of 60 diagrams
in which lepton lines and vertices of $\Lambda_4$
are modified by second-order radiative corrections.

We are now ready to classify the diagrams
into four (gauge-invariant) groups:

{\bf Group I.}  Second-order muon vertex diagrams containing lepton 
{\it v-p} loops $\Pi_2$, $\Pi_4$, $\Pi_{4(2)}$ and/or $\Pi_6$.
This group consists of 49 diagrams. 

{\bf Group II.}  Fourth-order proper vertex diagrams containing lepton 
{\it v-p} loops $\Pi_2$ and/or $\Pi_4$. 
This group consists of 90 diagrams. 

{\bf Group III.}  Sixth-order proper vertex diagrams containing a 
{\it v-p} loop $\Pi_2$.  This group consists of 150 diagrams.  

{\bf Group IV.}  Muon vertex diagrams containing 
an {\it l-l} subdiagram $\Lambda_4$ 
with additional 2nd-order radiative corrections,
or one of $\Lambda_4^{(2)}$ type.
This group consists of 180 diagrams.

All integrals of Groups I, II, and III have been evaluated
by numerical means.  Furthermore, some of them
have also been evaluated semi-analytically \cite{laporta3}.
Group IV integrals have thus far been evaluated only by numerical 
integration, but in two independent ways,
{\it Version A} and {\it Version B},
both in Feynman gauge.

The starting point of {\it Version A} is the RHS of Eq. (\ref{wtid2}).
The algebraic structure of integrals in {\it Version A}
is more complicated
than that of {\it Version B} but their codes are
substantially smaller in general than the latter.  
For Group I, however, there is no advantage of using {\it Version A}.
Thus this group is formulated in {\it Version B} only.

All integrals are generated from a small number of templates,
enabling us to make cross-checking of different diagrams, thereby 
reducing significantly possible programming errors.
More information on {\it Version A} and {\it Version B}
are given in Appendix
\ref{subsec:algebraic}.

Integrals thus obtained are divergent in general.
Since computers are not capable of handling divergence directly,
both ultraviolet (UV) and infrared (IR) divergences must be removed
beforehand.
We have introduced a two-step on-shell subtractive renormalization scheme,
in which the first step removes both UV and IR divergences
but does not give exact on-shell results.
This is done to circumvent the inconvenient feature of the
standard on-shell renormalization in which 
the renormalization terms do not remove and may even introduce
extra IR-divergent terms.
The second step yields the standard on-shell renormalization
result when summed over all diagrams.

The renormalization terms are generated in two ways:
One by reduction of the original integral
according to a well-defined power counting rule, and another
from scratch, both {\it analytically}.
This enables us to make extensive cross-checking between diagrams of
various types and different orders.
See Appendix \ref{renorm} for more details.

All integrals contributing to the $\alpha^4$ term 
are evaluated numerically
by the adaptive-iterative Monte-Carlo integration routine VEGAS \cite{lepage}.
The major source of numerical uncertainty is the difficulty
of accumulating a large number of good sampling points
that do not suffer from the {\it digit-deficiency} problem
caused by the round-off error.
For this purpose quadruple precision is required
in many cases.
Unfortunately, this slows down the computation
quite drastically.
The accuracy of these integrals is 
checked by comparison with those obtained by other means
whenever possible.
The results of our calculation are summarized in the 
following sections.
The reliability of these results,
which depends critically on the reliability
of the numerical integration routine VEGAS,
is discussed in Appendix \ref{ddproblem}.
Problems caused by non-statistical errors encountered in dealing with
VEGAS and their solution are discussed there in detail.

\section{Group I Diagrams}
\label{sec:group1}

\begin{figure}[ht]
\resizebox{9.5cm}{!}{\includegraphics{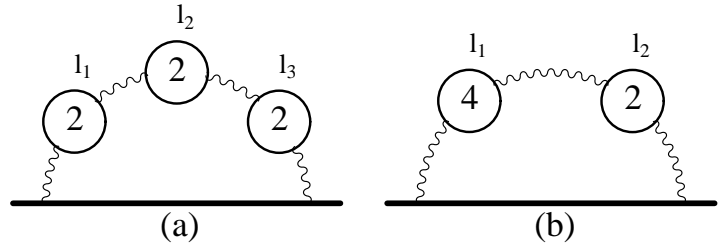}}
\caption{\label{vertex1} (a) Diagrams contributing to subgroup I(a).
(b) Diagrams contributing to subgroup I(b). 
Solid horizontal lines represent the muon in external magnetic field.
Numerals $``$2", $``$4" within solid circles refer to the 
proper renormalized {\it v-p} diagrams $\Pi_2$ and 
$\Pi_4$ , respectively.
Letters $l_1 , l_2 , l_3$ refer to electron or muon.
Seven and 18 Feynman diagrams contribute to I(a) and I(b),
respectively.}
\end{figure}

Group I diagrams can be classified further into
four gauge-invariant subgroups:

{\em Subgroup I(a)}.  
Diagrams obtained by inserting three $\Pi_2$'s
(of electron/muon loop) in a second-order muon vertex.  
Seven Feynman diagrams belong to this
subgroup.  See Fig. \ref{vertex1}(a).  

{\em Subgroup I(b)}.  
Diagrams obtained by inserting a $\Pi_2$ and a $\Pi_4$
in a second-order muon vertex.  
Eighteen Feynman diagrams belong to this subgroup.
See Fig. \ref{vertex1}(b).  

\begin{figure}[ht]
\resizebox{9.5cm}{!}{\includegraphics{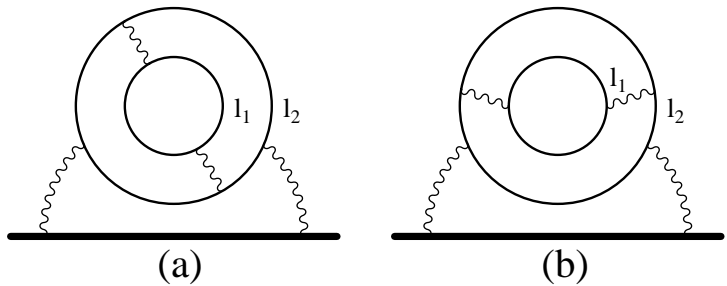}}
%
\caption{\label{vertex2} Diagrams contributing to subgroup I(c). $(l_1 , l_2 ) = (e, e), (e, \mu )$, or $(\mu ,e )$.
See FIG. 1 for notation.}  
\end{figure}

{\em Subgroup I(c)}.  
Diagrams containing $\Pi_{4(2)}$.
There are nine Feynman diagrams that belong to this subgroup. 
See Fig. \ref{vertex2}.  

\begin{figure}[ht]
\resizebox{9.5cm}{!}{\includegraphics{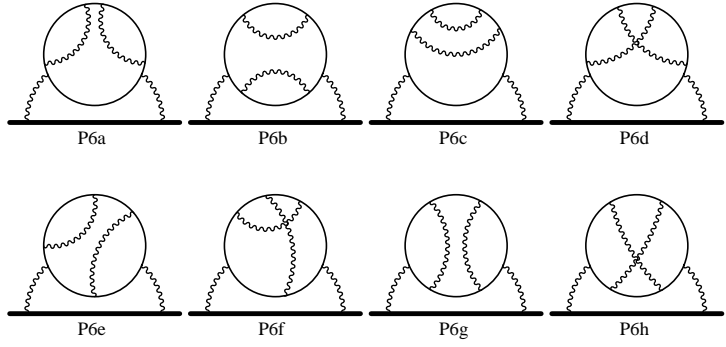}}
\caption{\label{vertex3} Eighth-order vertices of subgroup I(d) 
obtained by insertion of sixth-order (single electron loop) 
{\it v-p} diagram $\Pi_6$ in a second-order muon vertex.}  
\end{figure}

{\em Subgroup I(d)}.  
Diagrams obtained by insertion of $\Pi_6$
in a second-order muon vertex.  Fifteen
Feynman diagrams belong to this subgroup.  
Eight are shown in Fig. \ref{vertex3}.  
Diagrams $a, c,
d, e, f$ and the time-reversed diagram of $e$ have charge-conjugated
counterparts.

The evaluation of subgroups I(a) and I(b) is greatly
facilitated by the analytic formulas available for the second- and fourth-order 
spectral representations of the renormalized photon
propagators \cite{Kallen}.
The contribution to 
$a_{\mu}$ from the diagram obtained by sequential insertion of $m$ {\em k}-th
order electron and $n$ {\em l}-th order muon {\it v-p} loops into a
second-order muon vertex is reduced to a simple formula
\begin{equation}
a= \int_{0}^{1} dy(1-y) \left [ \int_{0}^{1} ds~{\frac{\rho_k (s)}{1~+~{\displaystyle {\frac{4}{1-s^2 }
{\frac{1-y}{y^2}} \left ( \frac{m_e}{m_{\mu} } \right )^2}} }} \right ]^m
\left [ \int_{0}^{1} dt~{\frac{\rho_l (t)}{1~+~{\displaystyle {\frac{4}{1-t^2} {\frac{1-y}{y^2}}}}}} \right ]^n ,           \label{a2mn}
\end{equation}
where $\rho_k$ is the {\em k}-th order photon spectral function.  
Exact $\rho_2$ and $\rho_4$ can be found in Ref. \cite{Kallen,KL1}.
An exact spectral function for $\Pi_{4(2)}$ and
an approximate one for $\Pi_6$ are also available \cite{hoang,broadhurst}.

The contribution of diagrams of Fig. \ref{vertex1} can be 
obtained by choosing $(k=2, m=3, n=0), (k=2, m=2,l=2, n=1), 
(k=2, m=1,l=2, n=2)$.
The latest numerical values obtained by
evaluating these integrals using VEGAS \cite{lepage} 
are listed in Table \ref{table1}, where the number of sampling points 
per iteration and the number of iterations are also listed.

Note that these diagrams need no additional renormalization.
Thus the renormalized amplitudes  
$a_{2,p2:3}^{(e,e,e)}$, etc. are given by
\begin{equation}
a_{2,p2:3}^{(e,e,e)} = M_{2,p2:3}^{(e,e,e)},~~~ {\rm etc.}
\label{scriptnotation}
\end{equation}
Note also that, to be consistent with notations used later,
$M_{2,p2:3}^{(e,e,e)}$, etc., should have been written as
$M_{2,p2:3}^{(\mu,e,e,e)}$, etc.
The first superscript $\mu$ is often (but not always)
suppressed for simplicity when there is no danger of confusion.

Adding up the first three rows of Table \ref{table1}, we obtain
 the total contribution of diagrams of subgroup I(a) 
\begin{equation}
a_{I(a)}^{(8)}~ =~7.745~140~(30)~~ .                    \label{a8Ia}
\end{equation}
This is about 40 times more precise than the earlier result \cite{km}.
Furthermore it is in excellent agreement with the results obtained by
an asymptotic expansion in $m_\mu / m_e$ \cite{laporta3}:
\begin{equation}
a_{I(a)}^{(8)} (asymp)~ =~7.745~136~8~(8) ,                    \label{a8Ia_asym}
\end{equation}
where the uncertainty comes only from the measurement of muon mass.

The contributions of Fig. \ref{vertex1}(b) for $(l_1 , l_2 ) = (e, e), (e, \mu )$, and $(\mu , e)$ can be written down in a similar fashion.  
The most recent results of numerical integration
by VEGAS  are listed in the last three rows of Table \ref{table1}.
These diagrams need no additional renormalization, too.
The sum of these results is the contribution of the
subgroup I(b) 
\begin{equation}
a_{I(b)}^{(8)} =~7.581~262~(50) .     \label{a8Ib}
\end{equation}
This again is in excellent agreement with the asymptotic expansion result \cite{laporta3} 
\begin{equation}
a_{I(b)}^{(8)} (asymp) =~7.581~275~5~(2) ,     \label{a8Ib_asym}
\end{equation}
where the uncertainty comes only from the muon mass.

\begin{table}
\renewcommand{\arraystretch}{0.80}
\begin{center}
\caption{Contributions of diagrams of Figs. \ref{vertex1}(a) and \ref{vertex1}(b).
$n_F$ is the number of Feynman diagrams represented by the integral.
These evaluations were carried out on $\alpha$ workstations in 2001.
\\
\label{table1}
} 
\begin{tabular}{lclrr}
\hline
\hline
~Integral~~& ~~~$n_F$~~~  &~Value (Error)~~~~~&~Sampling~per~& ~~~No. of~~~ \\ [.1cm]  
& &~~including $n_F$~&~iteration~~~~~& ~~iterations~  \\ [.1cm]   \hline
$M_{2,P2:3}^{(e,e,e)}$ &1&~7.223~077~(29)~&~~$1 \times 10^9$ \hspace{7mm}~~&100\hspace{4mm}~~\\ [.1cm]
$M_{2,P2:3}^{(\mu,e,e)}$&3 &~0.494~075~( 6)~&~~$1 \times 10^8$ \hspace{7mm}~~&100\hspace{4mm}~~\\ [.1cm]
$M_{2,P2:3}^{(\mu,\mu,e)}$&3 &~0.027~988~( 1)~&~~$1 \times 10^8$ \hspace{7mm}~~&60\hspace{4mm}~~\\ [.1cm]
$M_{2,P2,P4}^{(e,e)}$ &6&~7.127~996~(49)~&~~$1 \times 10^9$ \hspace{7mm}~~&100\hspace{4mm}~~\\ [.1cm]
$M_{2,P2,P4}^{(\mu,e)}$&6 &~0.119~601~( 3)~&~~$1 \times 10^8$ \hspace{7mm}~~&60\hspace{4mm}~~\\ [.1cm]
$M_{2,P2,P4}^{(e,\mu)}$&6 &~0.333~665~( 4)~&~~$1 \times 10^8$ \hspace{7mm}~~&60\hspace{4mm}~~\\ [.1cm]
\hline
\hline
\end{tabular}
\end{center}
\end{table}
\renewcommand{\arraystretch}{1}

In evaluating the contribution to $a_{\mu}$ from the 
9 Feynman diagrams of subgroup I(c) shown in FIG. \ref{vertex2}, 
our initial approach was to make use of the parametric integral
representation of the {\it v-p} term $\Pi_4^{(2)}$.
Following the two-step renormalization procedure,
these contributions can be written in the form
\cite{seeref}.
\begin{equation}
a_{I(c)}^{(8)}~ =~\sum_{(l_1 , l_2 )} a_{2,P4(P2)}^{(l_1 , l_2 )} ,
 \label{groupIc}
\end{equation}
where each term of
\begin{equation}
a_{2,P4(P2)}^{( l_1 , l_2 )}~=~ 
\Delta {M_{2,P4a(P2)}^{( l_1 , l_2 )}}
~+~ 2 \Delta {M_{2,P4b(P2)}^{( l_1 , l_2 )}}          
~-~ 2 \Delta B_{2,P2}^{(l_2 , l_1 )} M_{2,P2}^{( \mu , l_2 )} ,  
\label{a2p4p2}
\end{equation}
are finite integrals obtained by the 
$ {\bf K}_S $ renormalization procedure
described in Ref. \cite{KL1} and Appendix A.  The 
suffix $P2$ stands for the second-order {\it v-p} diagram
$\Pi_2$, $ P4 $ for the fourth-order {\it v-p}
diagram $\Pi_4$, while 
$P4 (P2)$ represents the diagram $\Pi_{4(2)}$.
$P4$ receives contributions from $ P_{4a} $ 
(vertex correction) and $ P_{4b} $ (lepton self-energy insertion),
$P4=P_{4a}+ 2 P_{4b}$.   

\renewcommand{\arraystretch}{0.80}
\begin{table}
\caption{ Contributions of diagrams of Fig. \ref{vertex2}.
$n_F$ is the number of Feynman diagrams represented by the integral.
Numerical work was carried out on $\alpha$ workstations during 2001.
\\
\label{table2}
} 
\begin{tabular}{lclrr}
\hline
\hline
~Integral~~& ~~~$n_F$~~~  &~Value (Error)~~~~~&~Sampling~per~& ~~~No. of~~~ \\ [.1cm]  
& &~~including $n_F$~&~iteration~~~~~& ~~iterations~  \\ [.1cm]   \hline
~~~$\Delta M_{2,P4a(P2)}^{(e,e)}$&1 &~0.597~477~1~(111)~&~~~~~~~$1 \times 10^9$ \hspace{4mm}~~&100\hspace{4mm}~~ \\ [.1cm]
~~~$\Delta M_{2,P4a(P2)}^{(e, \mu)}$&1 &~0.121~902~1~(58)~&~~~~~~~$1 \times 10^7$ \hspace{4mm}~~&100\hspace{4mm}~~ \\ [.1cm]
~~~$\Delta M_{2,P4a(P2)}^{(\mu, e)}$&1 &~0.021~017~1~(13)~&~~~~~~~$1 \times 10^7$ \hspace{4mm}~~&100\hspace{4mm}~~ \\ [.1cm]
~~~$\Delta M_{2,P4b(P2)}^{(e,e)}$&2 &~0.982~017~4~(109)~&~~~~~~~$1 \times 10^9$ \hspace{4mm}~~&100\hspace{4mm}~~ \\ [.1cm]
~~~$\Delta M_{2,P4b(P2)}^{(e, \mu)}$&2 &~0.099~244~1~(84)~&~~~~~~~$1 \times 10^7$ \hspace{4mm}~~&100\hspace{4mm}~~ \\ [.1cm]
~~~$\Delta M_{2,P4b(P2)}^{(\mu, e)}$&2 &~0.000~586~0~(~4)~&~~~~~~~$1 \times 10^7$ \hspace{4mm}~~&100\hspace{4mm}~~ \\ [.1cm]
\hline
\hline
\end{tabular}
\end{table}
\renewcommand{\arraystretch}{1}

The results of numerical evaluation of (\ref{a2p4p2}),
obtained by VEGAS,
are listed in Table \ref{table2}.  
Numerical values of 
lower-order Feynman integrals, in terms of which the residual renormalization
terms are expressed, are given in Table \ref{table3aux}.  From 
these tables  and (\ref{groupIc}) we obtain
\begin{equation}
a_{2,P4(P2)}^{(e,e)} =~1.440~744~(16)   ,          \label{a8Ic_ee}
\end{equation}
\begin{equation}
a_{2,P4(P2)}^{(e, \mu )} =~0.161~982~(11)  ,          \label{a8Ic_em}
\end{equation}
\begin{equation}
a_{2,P4(P2)}^{(\mu , e)} =~0.021~583~(2)  .          \label{a8Ic_me}
\end{equation}

\renewcommand{\arraystretch}{0.80}
\begin{table}
\caption{ Auxiliary integrals for Group I.
Some integrals are known exactly. Some are obtained by expansion in
$m_e /m_\mu$ to sufficiently high orders. Their uncertainties  come from
that of $m_e/m_\mu$ only.
The remaining integrals are obtained numerically by VEGAS.
Total sampling points are of order $10^{11}$.
\label{table3aux}
} 
\begin{tabular}{llll}
\hline
\hline
~~~Integral~~~  &~~~Value~(Error)~~~~& ~~~Integral~~~~   &~~~Value~(Error)~~~~~~  \\ [.1cm]   \hline
~$M_{2, P2}^{(\mu , e)}$~~~&~~1.094~258~282~7~(98)~& ~~$M_{2 ,P2}^{(\mu , \mu)}$~~~&~~0.015~687~421~$\cdots$ ~
~\\[.1cm]
~$M_{2 ,P2^*}^{(\mu , e)}$~~~&~-0.161~084~05 $\cdots$~ ~&~~~~&~~\\[.1cm]
~$\Delta B_{2}$~~~&~~0.75~~~~&~~$\Delta B_{2,P2}^{(e, e)}$~~&~~0.063~399~266~$\cdots$ \\[.1cm]
~$\Delta B_{2 ,P2}^{(\mu , e)}$~~~&~~1.885~732~6~(158)~~~~&~~$\Delta B_{2,P2}^{(e, \mu)}$~~&~~9.405~5~$\times~10^{-6}$~~\\[.1cm]
~$\Delta L_4$~~~&~~0.465~024~(12)~~~~&~~$\Delta B_4$~~&~-0.437~094~(21)~~\\
[.1cm]
~$\Delta \delta m_{4}$~~~&~~1.906~340~(22)~~~~&~~~&~~~\\
[.1cm]
\hline
\hline
\end{tabular}
\end{table}
\renewcommand{\arraystretch}{1}

\noindent
The new results (\ref{a8Ic_ee}) and (\ref{a8Ic_me})
confirm the old results but with a much higher precision.
For (\ref{a8Ic_em}) 
the agreement between the old and new values 
is rather poor.

About a decade ago the leading $log$ term
of $a_{2,P4(P2)}^{(e,e)}$
obtained by the renormalization group method \cite{faustov}
seemed to disagree with the numerical evaluation.
However, it was found \cite{KKO2} that this was caused by an improper use of the
asymptotic photon propagator obtained for  
massless QED in \cite{calmet}.  It is important to note that the 
asymptotic photon propagator for massless QED is not the same as one 
for massive QED as was proven explicitly in \cite{KKO2}.
Use of the correct photon propagator in
the renormalization group method leads to results
which agree very well with the numerical integration result
\cite{KKO, faustov2}.
This episode provides an explicit example 
of danger of confusing the
asymptotic behavior with the mass-less limit, which results in
different non-leading terms.

We obtained an independent check of (\ref{a8Ic_ee})
using an exact $\alpha^3$ spectral function for $\Pi_{4(2)}$
of Fig. \ref{vertex2}, which was 
derived \cite{spectralfn}
from the QCD spectral function obtained in \cite{hoang}.
Numerical integration using this spectral function gives
\begin{equation}
a_{2,P4(P2)}^{(e,e)} =~1.440~622~(173)  ,         \label{a8IcMN}
\end{equation}
for 100 million sampling points iterated 100 times
in quadruple precision.
This is in agreement 
with (\ref{a8Ic_ee})
to the fifth decimal point 
although their approaches are completely different.
Undoubtedly both (\ref{a8Ic_ee}) and (\ref{a8IcMN}) must be correct.

The best value of $a_{I(c)}^{(8)}$ is obtained by adding up
(\ref{a8Ic_ee}), (\ref{a8Ic_em}), and (\ref{a8Ic_me}):
\begin{equation}
a_{I(c)}^{(8)} =~1.624~308~(19)  .         \label{a8Ic}
\end{equation}

The contribution to $a_{\mu}$ from 15 diagrams of subgroup I(d) (see Fig. \ref{vertex3})
can be written as
\begin{equation}
a_{2,P6i}~=~\Delta M_{2,P6i}~+~{\rm {residual~renormalization~terms}}~, 
~~~~~(i~=~a,~ .~.~.~,~h).     \label{groupId}
\end{equation}
Divergence-free integrals $\Delta M_{2,P6i}$ are defined by (4.13)
of Ref. \cite{KL1}.  
Their numerical values (summed over the diagrams related by time-reversal and
charge-conjugation symmetries) 
are evaluated numerically by VEGAS
and listed in the third column of Table \ref{table2a}.

\renewcommand{\arraystretch}{0.80}
\begin{table}
\caption{ Contributions of diagrams of Fig. \ref{vertex3}.
$n_F$ is the number of Feynman diagrams represented by the integral.
$M_{2,P6e}$ was evaluated in 1998 on SP2 at Cornell Theory Center.
Others were evaluated in 1998 on Fujitsu VX at Nara Women's University, Japan.
\\
\label{table2a}
} 
\begin{tabular}{lclrr}
\hline
\hline
~Integral~~& ~~~$n_F$~~~  &~Value (Error)~~~~~&~Sampling~per~& ~~~No. of~~~ \\ [.1cm]  
& &~~including $n_F$~&~iteration~~~~~& ~~iterations~  \\ [.1cm]   \hline
~~~$\Delta M_{2,P6a}$&2 &~~5.676~002~(168)~&~~~~~$4 \times 10^8$ \hspace{4mm}~~&60\hspace{4mm}~~   \\ [.1cm]
~~~$\Delta M_{2,P6b}$&1 &~~3.058~301~(152)~&~~~~~$2 \times 10^8$ \hspace{4mm}~~&60\hspace{4mm}~~   \\ [.1cm]
~~~$\Delta M_{2,P6c}$&2 &~~1.483~501~(104)~&~~~~~$2 \times 10^8$ \hspace{4mm}~~&60\hspace{4mm}~~   \\ [.1cm]
~~~$\Delta M_{2,P6d}$&2 &~-3.127~282~(122)~&~~~~~$2 \times 10^8$ \hspace{4mm}~~&60\hspace{4mm}~~   \\ [.1cm]
~~~$\Delta M_{2,P6e}$&4 &~-0.073~885~(234)~&~~~~~$6 \times 10^8$ \hspace{4mm}~~&60\hspace{4mm}~~   \\ [.1cm]
~~~$\Delta M_{2,P6f}$&2 &~-4.064~113~(151)~&~~~~~$2 \times 10^8$ \hspace{4mm}~~&60\hspace{4mm}~~   \\ [.1cm]
~~~$\Delta M_{2,P6g}$&1 &~-0.247~237~(100)~&~~~~~$2 \times 10^8$ \hspace{4mm}~~&60\hspace{4mm}~~   \\ [.1cm]
~~~$\Delta M_{2,P6h}$&1 &~~2.838~657~( 74)~&~~~~~$2 \times 10^8$ \hspace{4mm}~~&60\hspace{4mm}~~   \\  [.1cm]
\hline
\hline
\end{tabular}
\end{table}
\renewcommand{\arraystretch}{1}

Summing up the contributions of diagrams $a$ to $h$ of Fig. \ref{vertex3},
we obtain the following expression:
\begin{eqnarray}   
a_{I(d)}^{(8)} &=~\sum_{i=a}^h \eta_i  \Delta M_{2,P6_i}
~-~4 \Delta B_2 \Delta M_{2,P4}^{( \mu ,e)}    \nonumber  \\
&+~5( \Delta B_2 )^2 M_{2,P2}^{( \mu ,e)}
~-~2( \Delta L_4~+~\Delta B_4) M_{2,P2}^{( \mu ,e)}   \nonumber  \\
&-~2 \Delta \delta m_4 M_{2,P2^*}^{( \mu ,e)} ,  
\label{a8Idform}
\end{eqnarray}
where
\begin{eqnarray} 
\Delta B_2  ~&=~{ \Delta}^{'} B_2 ~+~
{\Delta}^{'} L_2 ~=~ {\frac{3}{4}} ,    \nonumber  \\
{\Delta} M_{2,P4}^{( \mu ,e)} ~&=~\Delta M_{2,P4a}^{( \mu ,e)}~+~
2 \Delta M_{2,P4b}^{( \mu ,e)}  ,    \nonumber  \\
\Delta L_4 ~&=~\Delta L_{4x} ~+~2 \Delta L_{4c} +~\Delta L_{4l}
~+~2 \Delta L_{4s} ,    \nonumber  \\
\Delta B_4 ~&=~\Delta B_{4a} +~\Delta B_{4b} ,    \nonumber  \\
\Delta \delta m_4 ~&=~\Delta \delta m_{4a} +~\Delta \delta m_{4b} . 
\label{aux}
\end{eqnarray}
The quantities listed in (\ref{aux}) are defined in Ref. \cite{KL1}.  Their
numerical values are listed in Table \ref{table3aux}.  
The 1998 results of numerical integration of $\Delta M_{2,P6i}$
are listed in Table \ref{table2a}.  From 
the numerical values in Tables \ref{table3aux} and \ref{table2a} we obtain
the value reported previously \cite{kn2}:
\begin{equation}
a_{I(d)}^{(8)} =~-0.230~596~(416) .            
\label{a8Id}
\end{equation}

This deviates strongly from the old result $- 0.7945(202)$ \cite{km}.
The problem with \cite{km}
 was first pointed out in \cite{broadhurst} in which $a_{I(d)}^{(8)}$
was evaluated without the $\cal O$($m_e / m_{\mu}$) term
by a renormalization group method. Soon afterwards    
a Pad$\acute{e}$ approximant of the sixth-order photon spectral function
was used to evaluate the full correction \cite{baikov}:
\begin{equation}
a_{I(d)}^{(8)}(Pad\acute{e}) =~-0.230~362~(5) .            
\label{a8Id_Pade}
\end{equation}
Our new result (\ref{a8Id}) is in good agreement with (\ref{a8Id_Pade}).
The primary cause of the old discrepancy was traced to 
very poor statistics of the original evaluation \cite{KNO}.
Increase of statistics by two orders of magnitude
 improved the result to $-0.2415(19)$ \cite{kino2}.
However, the discrepancy with (\ref{a8Id_Pade}) was still non-negligible.
Finally, the problem was traced to round-off errors caused by
insufficient number of effective digits in real*8 arithmetic in carrying
out renormalization by numerical means 
\cite{kn2}.
This was resolved by going over to the real*16 arithmetic.
(See Appendix \ref{ddproblem} for further discussion on this point.) 

Note that the uncertainty in (\ref{a8Id_Pade}) 
may be an underestimate since it does not include the uncertainty
of the Pad\'{e} approximation itself.
However, it seems to be small compared with the quoted uncertainty \cite{kn2}.
In principle it is possible to prove or disprove it
by more numerical work.
However, it would require 6,000 times more computing time 
in order to match the precision 
of (\ref{a8Id_Pade}) 
achieved by the Pad\'{e} method. 
This is not only impractical but also pointless
since there is no need to improve the current precision further.

Collecting the results (\ref{a8Ia_asym}), (\ref{a8Ib_asym}), 
(\ref{a8Ic}) and (\ref{a8Id_Pade}), we find the best value of the
contribution to the muon anomaly from the 49 diagrams of group I: 
\begin{equation}
a_{~I}^{(8)}~=~16.720~359~(20)~~~ .             
\label{a8I}
\end{equation}

\section{Group II Diagrams}
\label{sec:group2}

Diagrams of this group are generated by inserting $\Pi_2$
and $\Pi_4$ in the photon lines of
fourth-order muon vertex diagrams.
Use of analytic expressions for the second- and fourth-order spectral
functions for the photon propagators
and time-reversal symmetry cuts down the number of 
independent integrals in {\it Version A}
from 90 to 11.

\begin{figure}[ht]
\resizebox{9.5cm}{!}{\includegraphics{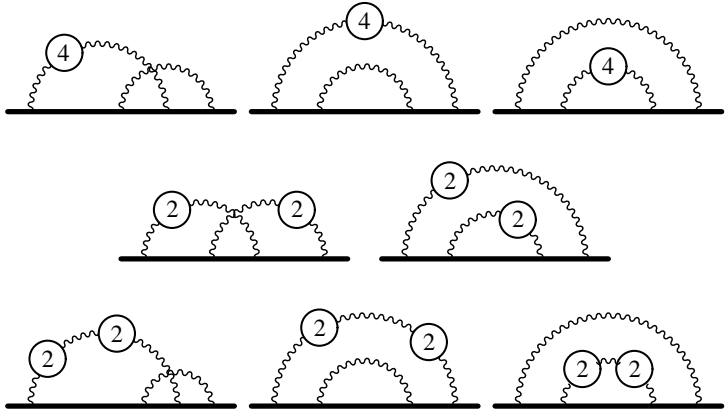}}
\vspace{0.5cm}
\caption{\label{vertex5} Eighth-order diagrams obtained from 
the fourth-order vertex diagrams by inserting vacuum-polarization loops
$\Pi_2$ and $\Pi_4$, which consist of either electron or muon loop.
}
\end{figure}

The contribution to $a_{\mu}$ arising from the set of vertex diagrams
represented by the $``$self-energy" diagrams of Fig. \ref{vertex5}
can be written in the form
\begin{equation}
a_{4,P_{\alpha}}~ =~\Delta M_{4,P_{\alpha}} +
~{\rm {residual ~renormalization~ terms}}  ,
\end{equation}
where $\Delta M_{4,P_{\alpha}}$ are finite integrals
obtained in the intermediate step of two-step
renormalization \cite{kino0}.  Their 
numerical values, obtained by VEGAS 
are listed in Table \ref{table4}.   The values of auxiliary integrals
needed to calculate the total contribution of group II diagrams are given in
Tables \ref{table3aux} and \ref{table4aux}.

Summing the contributions of diagrams of the first, second, 
and third rows of Fig. \ref{vertex5}, 
one obtains 
\begin{eqnarray} 
a_{4,P4} ~&=~2 \Delta M_{4a,P4}^{( \mu ,e)} +~
\Delta M_{4b,P {1^{'}} : 4}^{( \mu ,e)} ~+~\Delta M_{4b,P0:4}^{( \mu ,e)}~~~~   \nonumber  \\
 ~&-~{\Delta}  B_2 M_{2,P4}^{( \mu ,e)}
~-~{\Delta} B_{2,P4}^{( \mu ,e)} M_2 ,   
\label{P4form}
\end{eqnarray}
\begin{eqnarray} 
~~a_{4,P2,P2} ~&=~\Delta M_{4a,P2,P2}^{(e,e)}~+~
\Delta M_{4b,P1^{'} : 2 ,P0:2}^{(e,e)}     \nonumber  \\
~&-~ {\Delta} B_{2,P2}^{( \mu ,e)}~M_{2,P2}^{( \mu ,e)}~    \nonumber  \\
~&+~2 \Delta M_{4a,P2,P2}^{(e, \mu )}~+~
\Delta M_{4b,P1^{'} :2,P0:2}^{(e, \mu )}~+~
\Delta M_{4b,P1^{'} :2,P0:2}^{( \mu ,e)}    \nonumber  \\
~&-~ {\Delta} B_{2,P2}^{( \mu , \mu )} ~M_{2,P2}^{( \mu ,e)}
~-~ {\Delta} B_{2,P2}^{( \mu ,e)} ~M_{2,P2}^{( \mu , \mu )} ,  
\label{P2P2form}
\end{eqnarray}
\begin{eqnarray} 
a_{4,P2:2}~&=~2 \Delta M_{4a,P2:2}^{(e,e)}~+~
\Delta M_{4b,P {1^{'}} :2:2}^{(e,e)}~+~
\Delta M_{4b,P0:2:2}^{(e,e)}    \nonumber  \\
~&-~\Delta B_2 M_{2,P2:2}^{(e,e)}~-~
{\Delta} B_{2,P2:2}^{(e,e)}~M_2~    \nonumber  \\
~&+~4 \Delta M_{4a,P2:2}^{(e, \mu )}~+~
2 \Delta M_{4b,P1^{'} :2:2}^{(e, \mu )}~+~
2 \Delta M_{4b,P0:2:2}^{(e, \mu )}      \nonumber  \\
~&-~2 \Delta B_2 M_{2,P2:2}^{(e, \mu )} 
~-~2 {\Delta} B_{2,P2:2}^{(e, \mu )} ~M_2 ,   
\label{P22form}
\end{eqnarray}
respectively, 
where $M_{2,P4}^{( \mu ,e)}$ is equal to 
$\Delta M_{2,P4}^{( \mu ,e)} - 2 \Delta B_2 M_{2,P2}^{( \mu ,e)}$.  
The factor 2 in front of $\Delta M_{4a,...}$
accounts for equivalent diagrams obtained by time-reversal and
another factor 2 in front of 
$\Delta M_{4a,...}$ and $\Delta M_{4b,...}$
accounts for
interchange of electron and muon vacuum-polarization loops.
In contrast, the auxiliary integrals listed in Tables \ref{table3aux}
and \ref{table4aux} do not include multiplicity.
Following the convention adopted below Eq. (\ref{scriptnotation}),
the first superscript $\mu$ indicating the external muon line
is supressed for simplicity. For instance, 
$\Delta M_{4a,P2,P2}^{(\mu,e,e)}$ is written as $\Delta M_{4a,P2,P2}^{(e,e)}$.

\renewcommand{\arraystretch}{0.80}
\begin{table}
\caption{ Contributions of diagrams of Fig. \ref{vertex5}.
$n_F$ is the number of Feynman diagrams represented by the integral.
t.r. refers to time-reversed amplitude.
Numerical evaluation was carried out on $\alpha$ workstations in 2001.
\\
\label{table4}
} 
\begin{tabular}{lclrr}
\hline
\hline
~Integral~~& ~~~$n_F$~~~  &~Value (Error)~~~~~&~Sampling~per~& ~~~No. of~~~ \\ [.1cm]  
& &~~including $n_F$~&~iteration~~~~~& ~~iterations~  \\ [.1cm]   \hline
$\Delta M_{4a,P4}^{(\mu , e)}$ + t.r.&~18&~~2.047~838~(221)&$~~1 \times 10^9$ \hspace{8mm}~~&100\hspace{4mm}~~ \\ [.1cm]
$\Delta M_{4b,P0:4}^{(\mu , e)}$&&&&   \\ [.1mm]
~~~+$\Delta M_{4b,P1':4}^{(\mu , e)}$
&18&$-$2.486~595~(119)&$~~1 \times 10^9$ \hspace{8mm}~~&120\hspace{4mm}~~   \\ [.1cm]
$\Delta M_{4a,P2,P2}^{(e,e)}$ + t.r.&3&~~2.289~959~(144)&$~~1 \times 10^9$ \hspace{8mm}~~&100\hspace{4mm}~~  \\ [.1cm]
$\Delta M_{4a,P2,P2}^{(e, \mu )}$ + t.r.&6&~~$0.054~120~(~34)$&$~~1 \times 10^8$ \hspace{8mm}~~&100\hspace{4mm}~~  \\ [.1cm]
$\Delta M_{4b,P1':2,P0:2}^{(e,e)}$
&3&$-4.249~598~(~76)$&$~~1 \times 10^9$ \hspace{8mm}~~&100\hspace{4mm}~~ \\ [.1cm]
$\Delta M_{4b,P1':2,P0:2}^{(\mu ,e)}$&&&& \\ [.1cm]
~~~+$\Delta M_{4b,P1':2,P0:2}^{(e, \mu )}$&6
&$-0.485~108~(~14)$&$~~1 \times 10^8$ \hspace{8mm}~~&100\hspace{4mm}~~ \\ [.1cm]
$\Delta M_{4a,P2:2}^{(e, e )}$ + t.r.&6&$~~5.148~441~(377)$&$~~1 \times 10^9$ \hspace{8mm}~~&100\hspace{4mm}~~  \\ [.1cm]
$\Delta M_{4a,P2:2}^{(e, \mu )}$ + t.r.&12&$~~0.260~977~(103)$&$~~1 \times 10^8$ \hspace{8mm}~~&100\hspace{4mm}~~ \\ [.1cm]
$\Delta M_{4b,P0:2:2}^{(e, e )}$ &&&&   \\ [.1mm]
~~~+$\Delta M_{4b,P1':2:2}^{(e, e )}$
&6&$-8.633~608~(190)$&$~~1 \times 10^9$ \hspace{8mm}~~&100\hspace{4mm}~~ \\ [.1cm]
$\Delta M_{4b,P0:2:2}^{(e, \mu )}$ &&&&   \\ [.1mm]
~~~+$\Delta M_{4b,P1':2:2}^{(e, \mu )}$
&12&$-1.102~819~(~42)$&$~~1 \times 10^8$ \hspace{8mm}~~&100\hspace{4mm}~~   \\  [.1cm]
\hline
\hline
\end{tabular}
\end{table}
\renewcommand{\arraystretch}{1}

\renewcommand{\arraystretch}{0.80}
\begin{table}
\caption{ Auxiliary integrals for Group II.
Some integrals are known exactly. Some are obtained by expansion in
$m_e /m_\mu$ to necessary orders. Their uncertainties  come from
that of $m_e/m_\mu$ only.
Remaining integrals are obtained by VEGAS integration, with
total sampling points of order $10^{11}$.
\label{table4aux}
} 
\begin{tabular}{llll}
\hline
\hline
~~~Integral~~~  &~~~Value~(Error)~~~~& ~~~Integral~~~~   &~~~Value~(Error)~~~~~~  \\ [.1cm]   \hline
~$M_{2}$~~~&~~0.5~~~~&~~$M_{2,P4}^{(\mu ,e)}$~~&~~1.493~671~581~(8)~\\[.1cm]
~$ M_{2 ,P2:2}^{(e , e)}$~~~&~~2.718~655~7~(1)~~~~&~~$ M_{2,P2:2}^{(\mu ,e)}$~~&~~0.050~259~648~(1)~~\\[.1cm]
~$\Delta B_{2}$~~~&~~0.75~~~~&~~$\Delta B_{2,P2}^{(\mu ,\mu)}$~~&~~0.063~399~266~$\cdots$\\[.1cm]
~$\Delta B_{2 ,P2:2}^{(e , e)}$~~~&~~5.330~381~(61)~~~~&~~$\Delta B_{2,P2:2}^{(e, \mu)}$~~&~~0.236~018~(9)~~\\[.1cm]
~$\Delta B_{2 ,P4}^{(\mu , e)}$~~~&~~2.439~109~(53)~~~~&~~~~&~~~~\\[.1cm]
\hline
\hline
\end{tabular}
\end{table}
\renewcommand{\arraystretch}{1}

Substituting the data from Tables \ref{table3aux} and \ref{table4}
 into (\ref{P4form}), 
(\ref{P2P2form}), and (\ref{P22form}) we obtain
\begin{eqnarray} 
a_{4,P4} ~&=~-2.778~565~(253) ,      \nonumber  \\
a_{4,P2,P2} ~&=~-4.553~017~(~68)  ,  \nonumber  \\
a_{4,P2:2} &=~-9.342~599~(438) .     
\label{a8IInum}
\end{eqnarray}
Two of these terms were also evaluated in \cite{laporta3}
by an asymptotic expansion in $m_\mu /m_e$:
\begin{eqnarray} 
a_{4,P4} (asymp)~&=~-2.778~852~33~(5) ,       \nonumber  \\
a_{4,P2:2} (asymp) ~&=~-9.342~722~1~(5) .      
\label{a8IIasym}
\end{eqnarray}
They are in excellent agreement with the numerical integration results.

Combining the results (\ref{a8IIasym}) and the value of 
$a_{4,P2,P2}$ from (\ref{a8IInum}) 
we find the best value for the contribution of 90
diagrams of group II to be
\begin{equation}
a_{~II}^{(8)}~=~-16.674~591~(~68)  .     \label{a8II}
\end{equation}

\section{Group III Diagrams}
\label{sec:group3}

\begin{figure}[ht]
\resizebox{9.5cm}{!}{\includegraphics{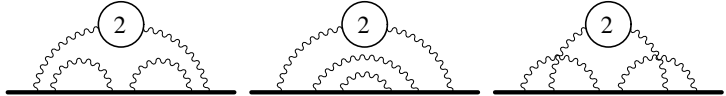}}
\vspace{0.5cm}
\caption{\label{vertex6} 
Typical eighth-order diagrams obtained by insertion of
a vacuum-polarization loop $\Pi_2$ in muon 
diagrams of the three-photon-exchange type.  Altogether there are
150 diagrams of this type.
}
\end{figure}

Diagrams belonging to this group are generated 
by inserting a second-order
vacuum-polarization loop  $\Pi_2$ in
the photon lines of sixth-order muon vertex
diagrams of the three-photon-exchange type.
Time-reversal invariance and use of the function 
${\rho}_2$ (see (\ref{a2mn})) for the
photon spectral function 
reduce the
number of independent integrals in {\it Version A} from 150 to 8.  
Some of these integrals are represented by the 
$``$self-energy" diagrams of Fig. \ref{vertex6}.

Let $M_{6 \alpha ,P}$ be the magnetic moment 
projection in {\it Version A}
of the set of 150 diagrams generated from a self-energy
diagram $ \alpha $ (=A through H) of Fig. \ref{vertex6} 
by insertion of $\Pi_2$ and
an external vertex.
The renormalized
contribution due to the group III diagrams can then 
be written as 
\begin{equation}
a_{III}^{(8)} ~=~\sum_{\alpha = A}^H { \eta}_{\alpha} a_{6 \alpha ,P2} ,
\end{equation}
where
\begin{equation}
a_{6 \alpha ,P2} ~=~\Delta M_{6 \alpha ,P2} ~+~ {\rm {residual~renormalization~terms}} .
\end{equation}
where all divergences have been projected out by ${\bf K}_S$ and
${\bf I}_R$ operations.  (See Ref. \cite{kino0}.)  

The latest numerical values
of Group III integrals are summarized in 
Table \ref{table5}.  
Numerical values of auxiliary integrals
needed in the renormalization scheme are listed in Tables \ref{table3aux},
\ref{table4aux} and \ref{table5aux}.
For comparison, the
results of
old calculation \cite{km} carried out in double precision
are listed in the last column of Table \ref{table5}. 
This is to examine the effect of {\it digit-deficiency} error.
In this case the effect is relatively mild because
the introduction of a vacuum-polarization loop tends to make
the integrand less sensitive to the singularity.

When summed over all the diagrams of group III, the UV- and IR-divergent pieces cancel
out and the total contribution to $a_{\mu}$ can be written as a sum of
finite pieces:
\begin{eqnarray} 
a_{III}^{(8)}~&=~ \sum_{ \alpha = A}^H
{\eta}_{\alpha} \Delta M_{6 \alpha ,P2}        \nonumber  \\
~&-~3 \Delta B_{2,P2}^{( \mu ,e)}  \Delta M_{4} -3 \Delta B_2  \Delta M_{4,P2}^{( \mu ,e)}              \nonumber  \\
~&+~( M_{2^{*} ,P2}^{( \mu ,e)} [I] - M_{2^* , P2}^{(\mu ,e)} )
\Delta \delta m_4
~+~(M_{ 2^{*}} [I] -M_{2^*} )  \Delta \delta m_{4,P2}^{( \mu , e)}
             \nonumber  \\
~&-~M_{2,P2}^{( \mu ,e)} [ \Delta B_4 + 2 \Delta L_4 ~-~ 2( \Delta B_2 )^2 ]                \nonumber  \\
~&-~M_2 ( \Delta B_{4, P2}^{(\mu, e)} + 2 \Delta L_{4,P2}^{( \mu ,e)} ~
~-~4 \Delta B_2 \Delta B_{2,P2}^{( \mu ,e)} ) .   
\label{aIII}
\end{eqnarray}

\renewcommand{\arraystretch}{0.80}
\begin{table}
\caption{ Contributions of diagrams of Fig. \ref{vertex6}.
$n_F$ is the number of Feynman diagrams represented by the integral.
This calculation was carried out in quadruple precision in 2001 - 2003
on $\alpha$ workstations to examine the influence of 
$digit$-$deficiency$ error in the 
calculation of \cite{km} carried out in double precision.
\\
\label{table5}
}
\begin{tabular}{lcrrrr}
\hline
\hline
~Integral~~& ~~~$n_F$~~~  &~Value (Error)~~~~~&~Sampling~per~& ~~~No. of~~~&~~~~Data from~~~ \\ [.1cm]  
& &~~including $n_F$~~~~~&~iteration~~~~~& ~~iterations &~~Ref. \cite{km}~~~~ \\ [.1cm]   \hline
$\Delta M_{6a,P2}$&15& $-$12.934~780~(1081)~&~~$4 \times 10^8$ \hspace{4mm}~~&100\hspace{4mm}~&$-$12.940~1~(130) \\[.1mm]
$\Delta M_{6b,P2}$&15&~~18.797~294~(1309)~&~~$4 \times 10^8$ \hspace{4mm}~~&140\hspace{4mm}~&18.797~0~(171) \\[.1mm]
$\Delta M_{6c,P2}$&15&~~~3.997~996~(1773)~&~~$4 \times 10^8$ \hspace{4mm}~~&100\hspace{4mm}~&4.000~7~(178) \\[.1mm]
$\Delta M_{6d,P2}$
&30&~~10.492~627~(1507)~&~~$8 \times 10^8$ \hspace{4mm}~~&111\hspace{4mm}~&10.494~0~(225) \\[.1mm]
$\Delta M_{6e,P2}$
&15&~~10.990~435~(~981)~&~~$4 \times 10^8$ \hspace{4mm}~~&119\hspace{4mm}~&11.000~1~(121) \\[.1mm]
$\Delta M_{6f,P2}$
&15&~~~5.652~451~(1503)~&~~$4 \times 10^8$ \hspace{4mm}~~&100\hspace{4mm}~&5.651~8~(166) \\[.1mm]
$\Delta M_{6g,P2}$
&30&~~19.747~805~(1558)~&~~$4 \times 10^8$ \hspace{4mm}~~&100\hspace{4mm}~&19.742~4~(172) \\[.1mm]
$\Delta M_{6h,P2}$
&15&$-$18.363~491~(1433)~&~~$4 \times 10^8$ \hspace{4mm}~~&100\hspace{4mm}~&$-$18.361~5~(141) \\ [.1cm]
\hline
\hline
\end{tabular}
\end{table}
\renewcommand{\arraystretch}{1}

\renewcommand{\arraystretch}{0.80}
\begin{table}
\caption{ Auxiliary integrals for Group III.
Some integrals are known exactly. Some are obtained by expansion in
$m_e /m_\mu$ to necessary orders. Their uncertainties  come from
that of $m_e/m_\mu$ only.
Remaining integrals are obtained by VEGAS integration, with
total sampling points of order $10^{11}$.
\\
\label{table5aux}
} 
\begin{tabular}{llll}
\hline
\hline
~~~Integral~~~  &~~~Value~(Error)~~~~& ~~~Integral~~~~   &~~~Value~(Error)~~~~~~  \\ [.1cm]   \hline
~$M_{2^*}$~~~&~~1.0~~~~&~ ~$M_{2^*}[I]$~~~&~~$-$1.0~~~\\[.1cm]
~$M_{2^*,P2}^{(\mu ,e)}$~~~&~~2.349~621~(~35)~~~~&~ ~$M_{2^*,P2}^{(\mu ,e)}[I]$~~~&~~$-$2.183~159~(95)~~\\[.1cm]
~$\Delta M_4$~~~&~~0.030~833~612~...~~~~&~~$\Delta M_{4 ,P2}^{(\mu , e)}$~~&~~$-$0.628~831~80~(2)~~\\[.1cm]
~$\Delta L_{4, P2}^{(\mu ,e)}$~~&~~3.118~868~(201)~   &
~$\Delta B_{4, P2}^{(\mu ,e)}$~~~&~~$-$3.427~615~(237)~~~ ~\\[.1cm]
~$\Delta \delta m_4$~~~&~~1.906~340~(22)~~~~&~~$\Delta \delta m_{4,P2}^{(\mu, e)}$~~&~11.151~387~(303)~~\\[.1cm]
\hline
\hline
\end{tabular}
\end{table}
\renewcommand{\arraystretch}{1}

Plugging the values listed in Tables 
\ref{table3aux}, \ref{table4aux} and \ref{table5aux}
in (\ref{aIII}), we obtain
\begin{equation}
a_{III}^{(8)} ~=~10.793~43~(414)  .  ~~    
  \label{a8III}
\end{equation}
The error in (\ref{a8III}) can be reduced easily if necessary.
The ration $a_{III}^{(8)}/\tilde{a}^{(6)}$, where $\tilde{a}^{(6)}$
is the value of sixth-order muon moment without closed lepton loop,
is about 11, which is not very far from the very crude
expectation $3K \sim 9$, where $K$ is from (\ref{Kdef}),
although such a comparison is more appropriate for
individual terms on Table \ref{table5} than their sum.
See Sec. VIII for further discussion of enhancement factor $K$.

\section{Group IV Diagrams}
\label{sec:group4}

Diagrams of this group can be divided into four subgroups:
IV(a), IV(b), IV(c), and IV(d).  Each subgroup
consists of two equivalent sets of diagrams related by charge conjugation 
(reversal of the direction of momentum flow in the loop of the light-by-light
scattering subdiagram).
Diagrams of subgroups IV(a), IV(b), and IV(c) are obtained
by modifying the sixth-order diagram which contains the 
light-by-light scattering subdiagram $\Lambda_4$, one of
whose external photon line represents the magnetic field. 
The magnetic moment contribution $M_{6LL}$ of this sixth-order diagram
is known analytically \cite{laporta2}, whose numerical value is
\begin{equation}
M_{6LL} =~20.947~924~34~(21)
\label{a6LL}
\end{equation}
when $\Lambda_4$ is an electron loop, and
the uncertainty is due to that of the muon mass only.

\begin{figure}[ht]
\resizebox{9.5cm}{!}{\includegraphics{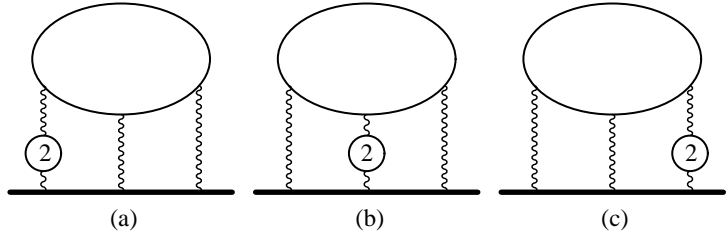}}
\vspace{.6cm}
\caption{\label{vertex7} Muon self-energy-like diagrams representing the external-vertex-summed integrals of subgroup IV(a). $(l_1 , l_2) = (e,e), (e, \mu)$, or $(\mu , e)$,
where $l_1,~l_2$ refer to the light-by-light scattering loop $\Lambda_4$ and vacuum-polarization loop $\Pi_2$, respectively. }  
\end{figure}

{\em Subgroup IV(a).}
Diagrams obtained by inserting a second-order
vacuum-polarization loop $\Pi_2$ in $M_{6LL}$.
They are all appropriate modifications of the integral
$M_{6LL,P2}$ defined by (2.4) of Ref. \cite{KL4}.  
Denote these integrals as
$M_{6LL,P2}^{( l_1 , l_2 )}$ where $( l_1 , l_2 )~=~
(e,e),~(e, \mu )$ or $( \mu ,e)$.  
This subgroup is comprised of 54 diagrams. 
They are generically represented by the
self-energy-like diagrams shown in Fig. \ref{vertex7}.

\begin{figure}[ht]
\resizebox{9.5cm}{!}{\includegraphics{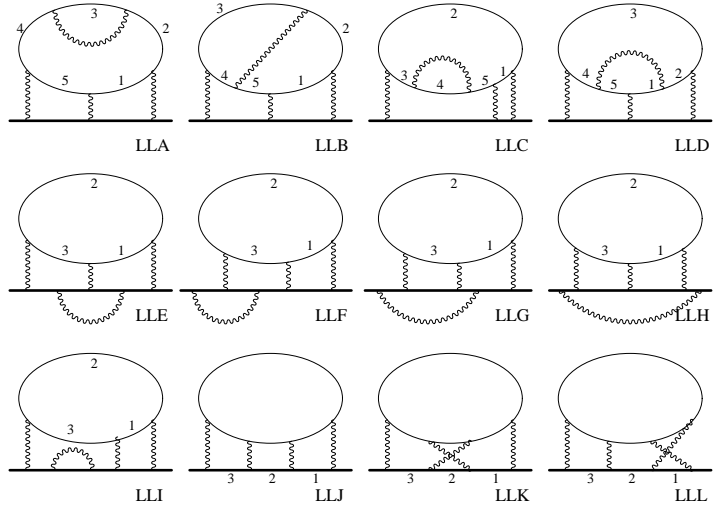}}
\vspace{.6cm}
\caption{\label{vertex8} Muon self-energy-like diagrams representing 
(external-vertex-summed) integrals of subgroup IV(b), IV(c), and IV(d).}  
\end{figure}

{\em Subgroup IV(b).}
Diagrams containing sixth-order light-by-light
scattering subdiagrams $\Lambda_6$.  Altogether, there are 60 diagrams of this
type.  Charge-conjugation and time-reversal 
symmetries and summation over external vertex insertions reduce
the number of independent integrals to 4 in {\it Version A}.  These integrals are
generically represented by the self-energy-like diagrams LLA, LLB, LLC and
LLD of Fig. \ref{vertex8}.

{\em Subgroup IV(c).}
Diagrams obtained by including 
second-order radiative corrections on
the muon line of $M_{6LL}$.
There are 48 diagrams that belong to this subgroup.  
Summation
over external vertex insertions and use of the interrelations
available due to charge-conjugation and time-reversal symmetries leave five
independent integrals in {\it Version A}.  
They are generically represented
by the self-energy-like diagrams 
LLE, LLF, LLG, LLH and LLI of Fig. \ref{vertex8}.

{\em Subgroup IV(d).}
Diagrams generated by inserting $\Lambda_4$
{\it internally} in fourth-order vertex
diagrams.  Diagrams of this type appear
for the first time in the eighth order.  
Charge-conjugation invariance and
summation over the external vertex insertion 
with the help of the Ward-Takahashi
identity lead us to three independent integrals in {\it Version A}.  
They are
represented by the diagrams LLJ, LLK and LLL of Fig. \ref{vertex8}.
No further discussion of this subgroup 
will be given in this paper since it
was treated in a separate paper \cite{kn1}.

In subgroups IV(a), IV(b), and IV(c) UV-divergences 
arising from the light-by-light scattering subdiagram
$\Lambda_4$, or more explicitly
$ {\Pi}^{ \nu \alpha \beta \gamma } (q, k_i , k_j , k_l )$,
can be taken care of by making use of the identity:
\begin{equation}
{\Pi}^{ \nu \alpha \beta \gamma } ( q, k_i , k_j , k_l ) ~=~
-~q_{\mu} \left [{\frac{\partial}{{ \partial q}_{\nu}} } {\Pi}^{ \mu \alpha \beta  
\gamma } ( q , k_i , k_j , k_l ) \right ] ,     \label{WTid}
\end{equation}
which follows from the Ward-Takahashi identity.
Namely, no explicit UV renormalization is needed
if one uses the RHS of (\ref{WTid}) instead of LHS
 and the fact that $\Sigma (p)$ of (\ref{wtid2})
vanishes by Furry's theorem.  
On the other hand, $\Sigma (p)$ is nonzero for subgroup IV(d) 
and the UV-divergence associated with the light-by-light
scattering subdiagram  $\Lambda_4$ must be regularized, e. g., 
by dimensional regularization.   For
these diagrams it is necessary to carry out explicit renormalization
of $\Lambda_4$ as well as that of the two sixth-order 
vertex subdiagrams containing $\Lambda_4$.  
See \cite{kn1} for a detailed discussion of renormalization
based on a combination of dimensional regularization and
Pauli-Villars regularization.

As was announced in Sec. \ref{sec:intro},
all diagrams of Group IV have now been evaluated in both {\it Version A}
and {\it Version B}.
In the following let us consider 
{\it Version A} and {\it Version B} separately
since renormalization is handled slightly differently in two cases.

\subsection{Version A}
\label{subsec:versa} 

The calculation of group IV(a) contribution is particularly simple.
This is because $M_{6LL}$ has been fully tested by comparison
with the analytic result \cite{laporta2}, and insertion
of the vacuum polarization term is straightforward.
One therefore finds that integrals $M_{6LL,P2}^{( l_1 , l_2 )}$ 
are all finite, which means
\begin{equation}
a_{6LL,P2}^{( l_1 , l_2 )} ~=~
M_{6LL,P2}^{( l_1 , l_2 )} ~=~
\Delta M_{6LL,P2}^{( l_1 , l_2 )} ,
\label{6LLP}
\end{equation}
where $l_1$, $l_2$ refer to the light-by-light scattering loop $\Lambda_4$
and vacuum-polarization loop $\Pi_2$, respectively.
Thus the contribution of subgroup IV(a) can be written as
\begin{equation}
a_{IV(a)}^{(8)} ~=~\sum_{( l_1 , l_2 )} \Delta M_{6LL,P2}^{( l_1 , l_2 )} ,  \label{a8IVa}
\end{equation}
where the individual terms are given in Table \ref{table6}.

\renewcommand{\arraystretch}{0.8}
\begin{table}
\caption{Contributions of diagrams of Fig. \ref{vertex7}.
$n_F$ is the number of Feynman diagrams represented by the integral.
The main term $\Delta M_{6LL,P2}^{(e,e)}$ was evaluated 
on $v1$ at Cornell Theory Center in 2001.
The rest was evaluated in 2001 on Condor cluster at University of Wisconsin.
\\
\label{table6}
}
\begin{tabular}{lcrrr}
\hline
\hline
~Integral~~& ~~~$n_F$~~~  &~Value (Error)~~~~~&~Sampling~per~& ~~~No. of~~~ \\ [.1cm]  
& &~~including $n_F$~&~iteration~~~~~& ~~iterations~  \\ [.1cm]   \hline
$\Delta M_{6LL,P2}^{(e,e)}$&18&$~116.759~183~(~292)$&~~$6 \times 10^{10}$ \hspace{4mm}~~&180\hspace{4mm}~ \\
$\Delta M_{6LL,P2}^{(e,\mu)}$&18&$~2.697~443~(~142)$&$1 \times 10^{8}$ \hspace{4mm}~~&110\hspace{4mm}~ \\
$\Delta M_{6LL,P2}^{(\mu ,e)}$&18&$~4.328~885~(~293)$&$1 \times 10^{9}$ \hspace{4mm}~~&100\hspace{4mm}~ \\
\hline
\hline
\end{tabular}
\end{table}
\renewcommand{\arraystretch}{1}

Let us denote magnetic projections of subgroups IV(b) and IV(c)
as $M_{8LL\alpha}$ where $\alpha = A,..,I$.
Relating the IR- and UV-divergent $M_{8LL \alpha }$ to the finite, numerically
calculable piece $\Delta M_{8LL \alpha}$ defined by the procedure of two-step
renormalization of Ref. \cite{KL4}, one can write 
the contributions of the diagrams of subgroups IV(b) and IV(c) as
\begin{equation}
a_{IV(b)}^{(8)} ~=~ \sum_{ \alpha = A }^D
{\eta}_{\alpha} \Delta M_{8LL \alpha } ~-~
3 \Delta B_2 M_{6LL} ,       \label{a8IVb}
\end{equation}
and
\begin{equation}
a_{IV(c)}^{(8)} =~\sum_{ \alpha = E }^I
{\eta}_{\alpha} \Delta M_{8LL \alpha } ~-~
2 \Delta B_2 M_{6LL} .          \label{a8IVc}
\end{equation}

\renewcommand{\arraystretch}{0.75}
\begin{table}
\caption{Contributions of diagrams of Fig. \ref{vertex8}
excluding LLJ, LLK, and LLL which were evaluated  separately in Ref. \cite{kn1}.
$n_F$ is the number of Feynman diagrams represented by the integral.
Some integrals are split
into two parts: $d$-$part$ is evaluated in real*8 and $q$-$part$
is evaluated in real*16. $a$-$part$ refers to the adjustable precision
method developed by \cite{sinkovits}.
The superscript *  indicates that indicated contributions 
were obtained by extrapolation from calculations
in which the edges of integration domain were
chopped off by 1.d-10.
See Appendix \ref{ddproblem} for details. 
Numerical work was carried out on SP3, velocity cluster, SP2, 
Condor cluster, and $\alpha$ workstations over several years.
The table lists only the latest of results obtained by various means.
\\
\label{table6a}
}
\begin{tabular}{lcrrr}
\hline
\hline
~Integral~~& ~~~$n_F$~~~  &~Value (Error)~~~~~&~Sampling~per~& ~~~No. of~~~ \\ [.1cm]  
& &~~including $n_F$~&~iteration~~~~~& ~~iterations~  \\ [.1cm]   \hline
$\Delta M_{8LLA}$ &10&$~~~~~~~52.063~459~(1497)$&$4 \times 10^8$&3300  \\
$\Delta M_{8LLB}$ &20&$~~~~~-75.014~508~(1838)$&~~&~   \\
~~~~{\it d-part}&&$~~-53.000~600~(~981)$&$1 \times 10^{10}$ &430   \\
~~~~{\it a-part*}&&$~-22.013~908~(1554)$&$4 \times 10^7$ &460   \\
$\Delta M_{8LLC}$ &20&$~~~~~107.488~810~(2811)$&~$4 \times 10^{8}$ &5900 \\
$\Delta M_{8LLD}$ &10&$~~~~~-37.824~352~(1137)$&~$1 \times 10^{10}$ &200 \\
$\Delta M_{8LLE}$ &6&$~~~~~-21.607~656~(1053)$&~~&~~   \\
~~~~{\it d-part}&&$~~-20.920~745~(~446)$&$1 \times 10^{10}$ &304 \\
~~~~{\it a-part}&&$~~-0.686~911~(~954)$&$2 \times 10^7$&280   \\
$\Delta M_{8LLF}$ &12&$~~~~~-75.765~816~(2341)$& $1 \times 10^{10}$ &~1000   \\
$\Delta M_{8LLG}$ &12&$~~-35.077~389~(1410)$&$1 \times 10^{10}$ &470   \\
$\Delta M_{8LLH}$ &6&$~~~~~~~54.025~704~(2411)$&~&~~  \\
~~~~{\it d-part}&&$~~51.820~951~( 889)$&$2 \times 10^{10}$ &470   \\
~~~~{\it q-part*}&&$~~2.204~753~(2241)$&$4 \times 10^7$&391   \\
$\Delta M_{8LLI}$ &12&$~~~~~~112.756~785~(2683)$&~$1 \times 10^{10}$ &450   \\
\\
\hline
\hline
\end{tabular}
\end{table}
\renewcommand{\arraystretch}{1}

Numerical integration of all terms 
contributing to $a_{IV}^{(8)}$ has been carried out using
VEGAS \cite{lepage}.  
The latest results 
for Groups IV(b) and IV(c) are listed in Table \ref{table6a}.  
The result for Group IV(d) had been handled separately \cite{kn1}.
In general, the major difficulty
in dealing with the diagrams of Groups IV(b) and IV(c)
 arises from the enormous size of
integrands (up to 5000 terms and 240 kilobytes of FORTRAN source code per integral) and the large number of integration
variables (up to 10).

Diagrams of Groups IV(b) and IV(c) have singular surfaces just outside
of the integration domain (unit cube)
at a distance of $\sim (m_e/m_\mu )^2 \sim 1/40,000$.
This makes the evaluation of their contributions to
$A_2^{(8)} (m_\mu /m_e )$ much more sensitive to the $d$-$d$ problem
compared with the evaluation of the same set of diagrams contributing to
the mass-independent $A_1^{(8)} $ whose singularity is far 
outside ($\sim 1$) of the
domain of integration.

Because of the $d$-$d$ problem intensified by
the proximity of the singularity,
all strategies discussed in Appendix \ref{ddproblem} had to be tried
to evaluate these integrals.

In most cases the first step is to make the integrand smoother
by {\it stretching} (see Appendix \ref{ddproblem}),
which is repeated several times until the integrand behaves
more gently.

Although {\it chopping} (see Appendix \ref{ddproblem}) was 
handy to obtain a rough estimate
quickly, we had to abandon it in the end
because extrapolation to $\delta = 0$ turned out to be too unreliable
in order to reach the desired precision.

Most integrals were then evaluated by {\it splitting} them into
two parts, one evaluated in real*8 and the other in real*16.
In some cases, however, even the part evaluated in real*16
suffered from severe $d$-$d$ problem, preventing us from
collecting large enough samplings for high statistics.
Analyzing this problem closely, we found that it is possible
to evaluate these integrals by the following procedure:
First try several iterations with a positive 
rescale parameter $\beta$ (typically $\beta = 0.5$)
until VEGAS begins to show strong sign of blowing up
due to the $d$-$d$ problem.
Then {\it  freeze} $\beta$ to 0 (see Appendix \ref{ddproblem}).
This may solve the problem in most cases.  If not,
try several iterations and see how rapidly the calculation runs into
the $d$-$d$ problem.
It turns out that it takes place very early 
if we chose too many sampling points $N_S$ per iteration.
This is because choosing large $N_S$ increases the chance of
hitting random numbers too close to the singularity within one iteration.
As a consequence the $d$-$d$ problem is likely to dominate
each iteration and makes it very difficult to collect
large enough number of good samplings.
We found that a better strategy is to reduce the size of $N_S$
to a moderate value and, instead, increase the number of iterations
$N_I$ substantially.
This is acceptable since, for $\beta = 0$
which means that the distribution function $\rho$ is
no longer changed from iteration to iteration, the final error
generated by VEGAS depends only on the product $N_S N_I$.

This strategy has been applied in particular to
the diagrams LLA and LLC, as is seen from Table \ref{table6a}.
Entries in Table \ref{table6a} are only the best of results obtained
by various methods discussed above.
They are consistent with each other despite their
diverse approaches.

One obtains from Tables \ref{table6} and \ref{table6a} 
the contributions of subgroup IV(a) and
the {\it Version A}  contributions
of subgroups IV(b) and IV(c):
\begin{eqnarray}   
a_{IV(a)}^{(8)} ~&=&~~123.785~51~(~44) ,   
  \nonumber  \\
a_{IV(b)}^{(8)} ~&=&~ - 0.419~42~(385) ,  
  \nonumber  \\
a_{IV(c)}^{(8)} ~&=&~~~~~2.909~41~(459) . 
\label{a8IV_A_num}
\end{eqnarray}   
%

\subsection{Version B}
\label{subsec:vb}

In {\it Version B} the magnetic moment projection 
is evaluated for each vertex diagram on
the LHS of (\ref{wtid2}).
It is convenient to denote these diagrams 
in terms of self-energy-like diagrams
of Fig. \ref{vertex8}, by attaching suffix $i$ to indicate
the lepton line in which an external magnetic field vertex
is inserted.
For instance, we obtain vertex diagrams LLA1, LLA2, ..., LLA5 from
the diagram LLA.

We will not discuss subgroup IV(a) here since  its {\it Version A}
has already been fully tested.
For subgroup IV(b) we find 
\begin{equation}
a_{IV(b)}^{(8)} ~=~ \sum_{ \alpha = A }^D 
\sum_{i=1}^5
{\eta}_{\alpha} \Delta M_{8LL \alpha i} ~-~
4 \Delta B_2 M_{6LL} ,       
\label{a8IVbB}
\end{equation}
instead of (\ref{a8IVb}).
Note that the last term of (\ref{a8IVbB}) is different
from that of (\ref{a8IVb}).
This is not an error.
It arises from difference in the definition
of $\Delta M$ terms.

Similarly, for subgroup IV(c) we obtain
\begin{equation}
a_{IV(c)}^{(8)} =~\sum_{ \alpha = E }^I \sum_{i=1}^3
{\eta}_{\alpha} \Delta M_{8LL \alpha i} ~-~
2 \Delta B_2 M_{6LL} .          \label{a8IVcB}
\end{equation}
The results of numerical evaluation are listed in Table \ref{tablexx}.
Precision of these calculations is still modest
but high enough to show the 
consistency with the calculation of {\it Version A}.
See the last column of Table \ref{tablexx} for comparison
of two Versions. 
Numerical work has been carried out with the same care
as that described for {\it Version A}.
The numerical calculation of $\Delta M_{8LLB}$ was particularly
difficult. 

One obtains from Table \ref{tablexx} the
values of $a_{IV(b)}^{(8)}$ and $a_{IV(c)}^{(8)}$
\begin{eqnarray}   
a_{IV(b)}^{(8)} ~&=&~ - 0.372~0~(168) ,  
  \nonumber  \\
a_{IV(c)}^{(8)} ~&=&~~~~2.876~3~(173) , 
\label{a8IV_numB}
\end{eqnarray}   
which are not inconsistent with those given in (\ref{a8IV_A_num}),
although much improvement is needed to become competitive
with the {\it Version A} results.

\begin{table}
\caption{Contribution of Group IV(b)
and Group IV(c) diagrams of Fig. \ref{vertex8}
evaluated in {\it Version B}.
Double precision is used for all calculations
which are carried out on Fujitsu VPP at RIKEN.   
Finite renormalization  terms $\Delta B_2 M_{6LLi}, i=1, 2, 3$, are 
needed for LLA and LLC, respectively, in order to compare  
them with the calculations in  {\it Version A}.   $M_{6LL(2+3)}
\equiv M_{6LL2} + M_{6LL3}$ is obtained
subtracting $M_{6LL1}$ from the known value of  
$M_{6LL}(\equiv M_{6LL1}+ M_{6LL2}+ M_{6LL3})$ given in (\ref{a6LL}).
\\
\label{tablexx}
}
\begin{tabular}{lcrrrr}
\hline
\hline
~Integral~~& ~~~$n_F$~~~  &~Value (Error)~~~~~&~Sampling~per~& ~~~No. of~~~
&~~~Difference ~~~ \\ [.1cm]  
& &~~including $n_F$~&~iteration~~~~~& ~~iterations~
&~~~ {\it Ver.A} - {\it Ver.B}   \\ [.1cm]   \hline
$\Delta M_{8LLA}$ &10&     52.080~79~(~731)     &&& $-$0.016~18~(~785)   \\
~~$\sum_{i=1}^5 {\eta}_A \Delta M_{8LLAi}$ 
& &   60.467~98~(~731) & $1 \times 10^{9}$  & 122~~~ &  \\
~~~~$-$$\Delta B_2 M_{6LL2}$ & &$-$8.387~20~(~~15)  & $1 \times 10^{10}$  & 280 ~~~&  \\ 
$\Delta M_{8LLB}$ &20& $-$74.999~66~(1060) & $1 \times 10^{9}$  & 544~~~ &$-$0.014~83~(1076)  \\
$\Delta M_{8LLC}$ &20&    107.503~69~(~877)  &&    &$-$0.012~44~(~955)    \\  
~~$\sum_{i=1}^5 {\eta}_C \Delta M_{8LLCi}$ 
 & &    114.827~43~(~876)  & $1 \times 10^{9}$  & 357 ~~~& \\
~~~~$-$$\Delta B_2 M_{6LL(1+3)}$ & &$-$7.323~75~(~~15)      &  &  &      \\ 
$\Delta M_{8LLD}$ &10&    $-$37.823~98~(~580)     & $1 \times 10^{9}$  & 120~~~ &$-$0.000~37~(~591) \\
$\Delta M_{8LLE}$ &6&   $-$21.611~47~(~562)    &$1 \times 10^{9}$ &120~~~  &$$+0.003~87~(~572)    \\
$\Delta M_{8LLF}$ &12&    $-$75.778~67~(~855)   &$1 \times 10^{9}$ &431~~~  &$$+0.014~28~(~921)   \\
$\Delta M_{8LLG}$ &12&    $-$35.074~71~(~683)   &$1 \times 10^{9}$ &120~~~  &$-$0.002~68~(~697)   \\
$\Delta M_{8LLH}$ &6&   54.013~78~(~619)  &$1 \times 10^{9}$ &262~~~  &$+$0.011~90~(~664)    \\
$\Delta M_{8LLI}$ &12&   112.749~26~(1037)  &$1 \times 10^{9}$ &512~~~  &+0.007~52~(1071)   \\
\hline 
\hline
\end{tabular}
\end{table}

\subsection{Total contribution of Group IV}
\label{subsec:grp4contrib}

The contribution of subgroup IV(a) is listed only in 
(\ref{a8IV_A_num}) since it was not evaluated in {\it Version B}.
The statistical combination of two versions of subgroups IV(b) and IV(c)
is dominated by {\it Version A} since {\it Version B} 
still does not have large statistics.
Only subgroup IV(d) has been evaluated in both versions
with comparable statistical weights \cite{kn1}.
Our best results for the
gauge-invariant subgroups of group IV can be summarized as
\begin{eqnarray}   
a_{IV(a)}^{(8)} ~&=&~~123.785~51~(~44) ,   
  \nonumber  \\
a_{IV(b)}^{(8)} ~&=&~ - 0.417~04~(375) ,  
  \nonumber  \\
a_{IV(c)}^{(8)} ~&=&~~~~~2.907~22~(444) , 
  \nonumber  \\
a_{IV(d)}^{(8)} ~&=&~-4.432~43~(~58) ,   
\label{a8IV_num}
\end{eqnarray}   
where $a_{IV(b)}^{(8)}$, $a_{IV(c)}^{(8)}$, and $a_{IV(d)}^{(8)}$ are statistical combinations
of {\it Version A} and {\it Version B}.

Summing up these terms one find that the contribution from
all 180 diagrams of group IV is given by
\begin{equation}
a_{IV}^{(8)} ~=~121.843~1~(59) .   \label{a8IV}
\end{equation}
Finally, combining (\ref{a8I}) with (\ref{a8II}), (\ref{a8III}) 
and (\ref{a8IV}), one obtains the value given in (\ref{newa28mue}).

\section{
Evaluation of
$A_3^{(8)} (m_\mu / m_e , m_\mu /m_\tau )$
}
\label{subsec:others}

There  is a small contribution to $a_\mu$ from the three-mass term
$A_3^{(8)} (m_\mu / m_e , m_\mu /m_\tau )$
which arises from 102 diagrams containing at least two closed fermion loops,
of {\it v-p} and/or {\it l-l} type.
The contribution of 30 diagrams analogous to those of
Fig. \ref{vertex1} and Fig. \ref{vertex2} is
\begin{equation}
A_{3I}^{(8)} ( m_{\mu} / m_e , m_{\mu} / m_{\tau} ) ~=~0.007~630~(~1).   
\label{a3I8}
\end{equation}
The contribution of 36 diagrams related to those of
Fig. \ref{vertex5} is
\begin{equation}
A_{3II}^{(8)} ( m_{\mu} / m_e , m_{\mu} / m_{\tau} ) ~=-0.053~818~(37).   
\label{a3II8}
\end{equation}
The contribution of 36 diagrams analogous to those of
Fig. \ref{vertex7} is
\begin{equation}
A_{3IV}^{(8)} ( m_{\mu} / m_e , m_{\mu} / m_{\tau} ) ~=~0.083~782~(75).   
\label{a3IV8}
\end{equation}

Summation of these results leads to the value given in (\ref{muetau}).
The value 0.079 (3) quoted in \cite{km} 
is in rough agreement with (\ref{a3IV8}).
In \cite{km} it was assumed that
the only nontrivial contribution to the eighth-order term 
arises from a muon vertex that contains an electron 
light-by-light scattering subdiagram and 
a tau vacuum-polarization loop and another 
in which the roles of electron and tau are interchanged. 
(See Fig. \ref{vertex7} with $(l_1 , l_2 ) = ( e, \tau ), (\tau , e)$.)  
It did not include the contributions 
(\ref{a3I8}) and (\ref{a3II8}).
Our new calculation shows that this assumption
was not justified.
This is presumably because the mechanism that makes $M_{6LL}$ 
enhanced (see discussion in Appendix \ref{renorm})
does not work if the momenta of photons exchanged between muon and
electron are not very small.

Another term of order $\alpha^4$ is 
$A_2^{(8)} (m_\mu / m_\tau )$ which is calculable from 469 Feynman diagrams.
However, its contribution to $a_\mu$ is of the order 
$(m_\mu /m_\tau )^2 \ln (m_\tau/m_\mu) A_2^{(8)} (1 ) \sim 0.005$
so that it may be safely ignored without actual calculation.

\section{Discussion}
\label{sec:discussion}

The size of integrals belonging to Groups I and II is rather small.
Thus they have been evaluated using large number of sampling
points, achieving precision of 5 or more digits.
Furthermore, most of these integrals have been evaluated
in alternative ways, either analytic or semi-analytic.
The agreement between numerical and (semi) analytic calculations is
so precise that it leaves no room for questioning the results.

A similar comment applies to the integrals of Group III
and Group IV(a), which are obtained 
by insertion of a vacuum-polarization loop $\Pi_2$
in the corresponding sixth-order diagrams, which
have been fully tested against the analytic integration results.

For the integrals of Group IV(d) formulation in {\it Version B}
enabled us to discover an error in the {\it Version A}.  
After correcting the error,
we now have two independent calculations which give
the same results.
For the remaining diagrams, of Groups IV(b) and IV(c),
their structure had been tested 
extensively taking advantage of the fact that they have
in general vertex and/or self-energy subtraction terms,
which can be generated in two ways:  One from a well-defined
reduction procedure of the original unrenormalized integral, 
and another by construction of the renormalization terms from scratch,
{\it both analytically}.
The agreement of these two,
proof of which often requires nontrivial analytic work, 
 give a strong confirmation
of their structure and 
of the master program from which all eighth-order integrals
of individual diagrams is derived.
See Appendix \ref{renorm} for details.

In order to obtain a further and definitive check, however, 
we have constructed IV(b) and IV(c) in {\it Version B}, too.
As a consequence, we have integrands of Groups IV(b) and IV(c) 
in two versions.
Extensive numerical work
has shown that they are consistent with each other within
the error bars of computation.
This completes a comprehensive check of all diagrams 
contributing to $A_2^{(8)} (m_\mu / m_e )$
by more than one independent methods.

It is important to note that our classification of diagrams into groups
(and subgroups) ensures that each subgroup is a {\it gauge-invariant} set.
In fact, individual integrals are mostly not gauge-invariant
and also infrared-divergent even when the UV divergence is renormalized
away.  On the other hand, each gauge-invariant set is 
well-defined and relatively small due to strong cancelation
among its constituents.
This is dramatically demonstrated by $a_{IV(b)}$ and $a_{IV(c)}$
in (\ref{a8IV_num}),
which are of order 1, whereas their constituents are 
two orders of magnitude larger as is seen from Table \ref{table6a}.

Empirically it is known that each mass-independent
minimal gauge-invariant sets contributing to g-2 
(namely diagrams containing no $v-p$ loop or $l-l$ loop)
is of order one apart from a power of $\alpha /\pi$.
When $v-p$ and/or $l-l$ loops are inserted in such diagrams,
they may acquire enhancements due to $\ln (m_\mu/m_e )$ factor,
which is a consequence of charge renormalization in case
of $v-p$ loop insertion and mass singularity present
in the limit $m_e \rightarrow 0$ in the case of $l-l$ loop.
The size of {\it mass-dependent} terms contributing to
$A_2^{(8)} (m_\mu /m_e )$
may be understood semi-quantitatively from this observation.

Let us now apply this argument to
$M_{6LL,P2}^{(e,e)}$ in Table \ref{table6},
which is gauge-invariant and yet very large.
Its size is inherited from the large sixth-order term $M_{6LL}$.
The extraordinary size of $M_{6LL}$ 
is due to the presence of $\ln (m_\mu/m_e)$ term 
with a large coefficient (6.38(8)) as
was initially discovered by numerical integration \cite{aldins}.
It was noted then (unpublished) that it is numerically
close to $2\pi^2/3$, later verified analytically \cite{lautrup},
enabling us to write
\begin{equation}
M_{6LL} = \frac{2\pi^2}{3} \ln (m_\mu /m_e )  + \cdots.
\label{leadingM6LL}
\end{equation}
Since $M_{6LL}$ is UV-finite, the term $\ln m_\mu$ comes from the scale
set by the largest physical mass of the system, $m_\mu$.
The $\ln m_e$ term arises from the integration of the
momentum $k$ of the $l-l$ loop $\Lambda_4$
over the domain $D_1$ ($m_e < | k| < m_\mu$, $|p_i| \leq m_e$),
where $p_i, (i=1,2,3)$ are the momenta of photons exchanged between the electron
and the muon.
Other domains such as $D_3$(any $k$, $|p_i| > m_e $)
does not contribute to $\ln m_e$.

What makes $M_{6LL}$ really large, however, is the presence of 
the coefficient $\pi^2$ in (\ref{leadingM6LL}).
A physical interpretation for this fact was given by
Elkhovskii \cite{yeli} who pointed out that, 
in the sub-domain $D_2$ ($m_e < | k| < m_\mu$, $|p_i| << m_e$, or more precisely
$|p_i| \lesssim \alpha m_e$, $\alpha  \simeq 1/137$)
the muon is nearly at rest and 
the electron can be treated as a non-relativistic particle
in the field of the muon. 
One of the photons
is responsible for the hyperfine spin-spin interaction, and the
other two act essentially like a static Coulomb potential.
It is the integration over the Coulomb photon momenta that gives a factor
$i\pi $ each, contributing a factor $\pi^2 (\sim 10)$
in (\ref{leadingM6LL}).

Actually it is not easy to
maintain the nonrelativistic behavior of the electron
throughout the domain $D_1$ outside of $D_2$. 
This will result in the erosion of the enhancement factor $\pi^2$
in the domain $D_1 - D_2$, although the $\ln (m_\mu/m_e)$
behavior is still maintained.
Together with non-logarithmic contribution from other parts of
momentum space, the net effect is to reduce the contribution
of the leading term of (\ref{leadingM6LL}), which is about 35,
to about 21.  
This reduction may be expressed crudely by
choosing the fudge factor $\xi$ to be about 0.12 in
\begin{equation}
M_{6LL}^{(approx)}  = \frac{2\pi^2}{3} \ln (\xi m_\mu /m_e )  .
\label{approxM6LL}
\end{equation}

Next consider the effect of the
renormalized photon propagator: 
\begin{equation}
D_R^{\mu ,\nu} (q) = -i \frac{g_{\mu \nu}}{q^2 } d_R (q^2 /m_e^2 , \alpha )
+ \cdots ,
\label{ren.p.p}
\end{equation}
where, to order $\alpha$,
\begin{equation}
 d_R (q^2 /m_e^2 , \alpha ) = 1 + \frac{\alpha}{\pi}
\left [ \frac{1}{3} \ln (q^2 /m_e^2 ) - \frac{5}{9} + \cdots \right ] .
\label{d_R}
\end{equation}
When $D_R$ is inserted in $g-2$ diagrams,
the scale for the momentum $q$ is set by the muon mass.
This means that the leading term of the integral
containing a vacuum-polarization loop $\Pi_2$ such as
$M_{6LL,P2}^{(e,e)}$ may be written as
\begin{equation}
M_{6LL,P2}^{(e,e)} \simeq 3 K M_{6LL}^{(approx)} + {\rm terms~ linear~in}~ \ln (m_\mu /m_e ),
\label{leading}
\end{equation}
where  the factor 3 is the number of photon lines in which
a vacuum-polarization loop can be inserted, and
\begin{equation}
K \equiv \frac{2}{3} \ln (m_\mu /m_e ) - \frac{5}{9} \simeq 3 ,
\label{Kdef}
\end{equation}
provided that $q^2$ is replaced by $m_\mu^2$ in (\ref{d_R}).
The combination of these factors 
is responsible for the large size of the leading term of
(\ref{leading}):
\begin{equation}
3 \times ((2/3) \ln (m_\mu /m_e ) -5/9 ) \times 20 \simeq 180,
\label{leadingM6LLP2}
\end{equation}
which is 1.5 times larger than the  calculated value of
$M_{6LL,P2}^{(e,e)}$ listed in Table \ref{table6},
pretty close for a crude approximation.
If the argument given below Eq. (\ref{approxM6LL})
is applicable to the photon momenta, we would obtain
$K \sim 1.7$ and (\ref{leadingM6LLP2}) would become $\sim 100$.
Both estimates would be acceptable as rough measure.

It is important to note that
$M_{6LL,P2}^{(e,e)}$
is not only very large but also its value is known very precisely
because it is obtained from the exactly known $M_{6LL}$
by a well-understood vacuum-polarization insertion  procedure.
This means that the value of the term $A_2^{(8)} (m_\mu /m_e )$
is determined primarily by $a_{IV(a)}$ while its uncertainty
comes mostly from $a_{IV(b)}$, $a_{IV(c)}$, and $a_{IV(d)}$.
This is why the value of $A_2^{(8)} (m_\mu /m_e )$ did not change
much even if $a_{IV(d)}$ suffered from a program error.

These arguments may also be applied in identifying the leading terms 
of  the tenth-order contribution $A_2^{(10)} (m_\mu /m_e )$
\cite{knX}.

\begin{acknowledgments}

The part of the material presented here by T. K.
is based upon work supported by the National
Science Foundation under Grant No. PHY-0098631.
M. N. is supported in part by the Ministry of Education, Science, Sports, 
and Culture, Grant-in-Aid for Scientific Research (c), 15540303, 2003.
T. K. thanks RIKEN, Japan, for the hospitality extended to him
where a part of this work has been carried out.
We thank A. Kataev for useful communications.
Thanks are due to  J. Zollweg, T. Tannenbaum,
and J. Ballard for assistance in various phases of computation.

The numerical work has been
carried out over several years on a number of computers.
A part of numerical work was conducted at
the Cornell Theory Center 
using the resources of the Cornell University,
New York State,
the National Center for Research Resources at 
the National Institute of Health,
the National Science Foundation, the Defense Department
Modernization Program, the United States Department of Agriculture,
and the corporate partners. 
Another part of numerical work was supported by 
NSF Cooperative agreement ACI-9619020 through computing resources
provided by the National Partnership for Advanced
Computational Infrastructure at the San Diego Supercomputer Center,
which enabled us to have an access to the Blue Horizon
at the San Diego Supercomputer Center,
the IBM SP at the University of Michigan, and the Condor Flock
at the University of Wisconsin.

M. N. also thanks for various computational resources provided by 
the Computer Center of Nara Women's University, RIKEN Supercomputing
System, and Matsuo Foundation.

\end{acknowledgments}

\appendix
\section{~~Elimination of algebraic error of integrals}
\label{description}

Most eighth-order integrals, including those of Group IV, are huge.
A systematic approach is required to make sure
that they are free from algebraic error and have 
forms suitable for numerical integration.
To achieve this we adopted the following procedure:

(a)  Carry out momentum integration of Feynman diagrams
and convert them into integrals over Feynman parameters
using algebraic manipulation program such as FORM \cite{form}. 
This step is fully analytic.
Conversion of all integrals of a gauge-invariant set is performed 
using a common {\it template}
by permuting tensor indices of photon propagators. 

(b)  Integrals thus obtained are divergent in general.
Since computers are not capable of handling divergence directly,
both ultraviolet (UV) and infrared (IR) singularities must be removed
from the integrand before integration is performed.
In the subtractive on-shell renormalization 
\cite{dimensionalreg}
of the $n$-th order diagram $M_n$, 
the renormalization  term involving a $m$-th
order vertex renormalization constant $L_m$ is of the form
$-L_m M_{n-m}$, where $M_{n-m}$ is the {\it g-2} term of order $n-m$.
The subtraction procedure described in textbooks is not suitable
for numerical integration, however, 
since it does not make the integrand of $M_n - L_m M_{n-m}$ finite
throughout the domain of integration,
as long as $L_m$ is just a numerical constant. 

The first step to achieve a point-wise cancelation 
is to express $L_m$ as a parametric integral
and combine it with the parametric integral of $M_{n-m}$ 
using a generalization of Feynman's formula 
\begin{equation}
\frac{1}{AB} = \int \frac{\delta (1-z_1-z_2)dz_1 dz_2}{(z_1 A + z_2 B)^2},
~~~z_1, z_2 \geq 0.
\label{feynmanformula} 
\end{equation}
The domain of the combined integral $L_m \bigotimes M_{n-m}$
may then be chosen to be identical with that of $M_n$.
Unfortunately, the integral is found to be intractable
if we want to treat $L_m$ as a whole.
However, if it is split as 
\begin{equation}
L_m =  L_m^{(UV)} + L_m^{(IR)} + L_m^{(finite)},
\end{equation}
where  the UV-divergent part $L_m^{(UV)}$ is identified by the highest power of $U$, 
the IR divergent part $L_m^{(IR)}$ by the highest power of $V$, 
$L_m^{(UV)} \bigotimes M_{n-m}$ is found to have 
a term-by-term correspondence with UV divergent terms of the
original (mother) integral $M_n$, 
and UV divergences of $M_n$ and $L_m^{(UV)} \bigotimes M_{n-m}$ cancel
each other before (not after) integration is performed.
(See (\ref{def0}) for the definitions of $U$ and $V$.)
IR divergence can be handled in a similar way.
This point-wise subtraction is crucial for the success of our
renormalization program on a computer.

(c)  In practice it is easier to start from the UV-divergent terms
of the mother integral $M_n$, which can be readily identified
by a power counting at the singularity,
and construct the subtraction term 
$L_m^{(UV)} \bigotimes M_{n-m}$ taking advantage of the term-by-term correspondence
described above.
This procedure, formalized as K-operation (see (\ref{vertexrenorm})
and (\ref{selfenergysubtr})) can be applied to all orders.
Further advantage of the K-operation is that it can be readily implemented
in the FORM program that generates the integrand of $M_n$.
It is easy to confirm that 
the UV singularities of $L_m^{(UV)} \bigotimes  M_{n-m}$ and $M_n$
cancel out exactly 
by numerically evaluating the mother and daughter
integrands at a sequence of points converging to the singular point.

(d)  By construction the
daughter integral factorizes {\it analytically} into a product of 
the {\it divergent part} of a renormalization
constant and a magnetic moment of lower order.
Remaining parts of renormalization constants, such as $L_m^{(finite)}$,
are summed over all diagrams afterwards.
The result is a convergent integral of lower order,
which is easy to evaluate numerically.
In other words, our renormalization proceeds in two steps. 
But, of course, it is identical with the standard on-shell renormalization.

This procedure is designed to enable us to
obtain extensive cross-checking at every step.
To make this paper as self-contained as possible, 
let us describe them in some detail,
although they can all be found in previous papers \cite{kino0}.

\subsection{Construction of Feynman-parametric integral}
\label{subsec:algebraic}

Let $G$ be a $2n$-th order proper lepton vertex of QED,
which describes the scattering of an incoming lepton of
momentum $\pslash -\qslash/2$  by an external magnetic field
into an outgoing lepton
of momentum $\pslash + \qslash /2$, where both leptons are on the mass shell.
$G$ consists of $2n$ lepton propagators and $n$ photon propagators
of the form (in Feynman gauge)
\begin{equation}
 \frac{\kslash_i+\qslash_i +m_i}{(k_i + q_i )^2 - m_i^2},
~~~~~~ \frac{g_{\mu \nu}}{(k_i + q_i )^2 - m_i^2 },
\label{propagators}
\end{equation}
besides factors describing the interaction, spinors, etc.
Here $k_i $ is a linear combination of the loop momenta flowing
through the line $i$.
$q_i$ is a linear combination of external momenta.
$m_i$ is the mass associated with the line $i$, which is temporarily
distinguished from each other for technical reason.
All these factors are combined
to form a proper vertex part, which is integrated over $n$
loop momenta.

The first step of momentum integration is to replace 
$\kslash_i + \qslash_i$ in the numerator 
of (\ref{propagators}) by an operator \cite{karplus}
\begin{equation}
D_i^\mu \equiv \frac{1}{2} \int_{m_i^2}^{\infty} dm_i^2 \frac{\partial}{\partial q_{i\mu}}
\label{diff-int-op}
\end{equation}
for each lepton line $i$.
Since $D_i^{\mu}$ does not depend on $k_i$ explicitly, 
the numerator (turned into an operator now) 
can be pulled in front of momentum integration.
The integrand then becomes just a product of denominators,
which can be combined into one big function 
with the help of Feynman parameters $z_1, z_2, \ldots, z_N~ (N=3n)$,
assigned to respective propagators.
Momentum integration can now be carried out exactly.
$D_i^{\mu}$ can then be pulled back inside $z$ integration.
Omitting the factor $(\alpha /\pi )^n$ for simplicity
the result can be expressed in the form
\begin{equation}
{\bf \Gamma}_\nu^{(2n)} =
\left ( \frac{-1}{4} \right )^n (n-1)!~\int ~{\bf F}_\nu \frac{(dz)_G}{U^2 V^n},
\label{vertexdgrm}
\end{equation}
where
\begin{eqnarray}
(dz)_G &=& \delta (1- \sum_{i=1}^{N} z_i ) \prod_{i=1}^N dz_i, 
~~~V= \sum_{i=1}^N z_i (m_i^2 -q_i \cdot Q_i^{'}),   \nonumber  \\
Q_i^{'\mu} &=& - \frac{1}{U} \sum_{j=1}^N q_j^\mu z_j B_{ij}^{'},~~~ 
B_{ij}^{'} = B_{ij} - \delta_{ij} \frac{U}{z_j} .
\label{def0}
\end{eqnarray}
$U$ and $B_{ij}$ are homogeneous forms of degree $n$ 
and $n-1$ in $z_1, \ldots , z_N$, respectively.
(See \cite{kino0} for explicit definitions.)

The operator ${\bf F}_\nu$ has the form
\begin{eqnarray}
{\bf F}_\nu &\propto&
\gamma^{\alpha_1}
(\rlap{\hspace{0.07cm}/}{D}_1+m_1)
\gamma^{\alpha_2}
(\rlap{\hspace{0.07cm}/}{D}_2+m_2)
\ldots    \gamma_\nu \ldots                 
\gamma^{\alpha_{(2n-1)}}
(\rlap{\hspace{0.07cm}/}{D}_{(2n)}+m_{(2n)})
\gamma^{\alpha_{(2n)}}.
\label{numfactor}
\end{eqnarray}
If $G$ has closed lepton loops ${\bf F}_\nu$ contains some
trace operations, too.
The action of ${\bf F}^\nu$ on $1/V^n$ in (\ref{vertexdgrm})
produces terms of the form
\begin{equation}
{\bf F}^{\nu} \frac{1}{U^2 V^n} = \frac{F_0^{\nu}}{U^2 V^n}
+ \frac{F_1^{\nu}}{U^3 V^{n-1}}
+ \frac{F_2^{\nu}}{U^4 V^{n-2}} + \ldots,
\end{equation}
where the subscript $k$ of $F_k^{\nu}$ stands for the number of contractions.
By contraction we mean picking a pair $\Dslash_i + m_i$,
$\Dslash_j + m_j$ from ${\bf F}^\nu$, making the substitution
\begin{equation}
(\Dslash_i + m_i , ~~~\Dslash_j + m_j)  \Longrightarrow (\gamma^\mu ,~~~ \gamma_\mu) , 
\end{equation}
multiplying the result with $-\frac{1}{2} B_{ij}$,
and summing them over all distinct pairs.
Uncontracted $D_i$ are replaced by $Q_i^{'}$.
For $k \geq 1$, $F_k^{\nu}$ includes an overall factor $(n-1)^{-1} \cdots (n-k)^{-1}$.

In our problem it is convenient to use,
instead of the vector $Q_i^{'\mu}$ itself,
 a scalar function extracted from $Q_i^{'\mu}$.
Suppose $p_\mu-q_\mu/2$ (external lepton momentum) enters the graph $G$ 
at a point $A$, $q_\mu$ enters at a point $C$ 
(which is the magnetic field vertex),
 and $p_\mu + q_\mu/2$ leaves at a point $B$.
Then we may write
\begin{equation}
Q_i^{'\mu} = A_i^{(AC)} (p^\mu- q^\mu/2) + A_i^{(CB)} (p^\mu + q^\mu/2). 
\end{equation}
After a little manipulation we find, for example,
\begin{equation}
A_i^{(AB)} \equiv A_i^{(AC)} + A_i^{(CB)}
 = - \frac{1}{U} \sum_{j=1}^N \eta_{jP} (z_j B_{ji}- \delta_{ij} U) ,
~~~~~~~~~~P=P(AB).
\label{defAi}
\end{equation}
associated with the path $P=P(AB)$ of the external momentum $p$,
which is any self-non-intersecting path starting at $A$ 
and ending at $B$, and $\eta_{jP} = (1, -1,0)$ according to whether 
the line $j$ lies (along, against, outside of) the path $P$.
$A_i^{(AC)}$, $A_i^{(CB)}$, etc., are
called ``scalar currents" since they satisfy an
analog of Kirchhoff's laws for electric currents
when the diagram $G$ is regarded as an electric network
in which Feynman parameter $z_i$ plays the role of resistance \cite{bjorken}.
$B_{ij}$ satisfies Kirchhoff's laws, too.
$U$ and $B_{ij}$ depend only on the topology of the graph $G$ and not 
on whether the line is fermion or boson.
They can be constructed easily from their definitions or by recursive
relations starting from the one-loop case.
We have written MAPLE and FORM programs to compute them algebraically
for an arbitrary diagram.
Once $U$ and $B_{ij}$ are known, $A_i \equiv A_i^{(AB)}$ can be constructed 
by (\ref{defAi}).
For further details see \cite{kino0}.

The magnetic moment projection of ${\bf \Gamma}_\nu^{(2n)} $
of the muon in {\it Version B} is given by
\begin{equation}
M_G^{(2n)B} = \lim_{q=0} Tr[P^\nu (p,q) \Gamma_\nu^{(2n)}],
\label{magmoment}
\end{equation}
where $(p + q/2 )^2 = (p - q/2 )^2 = m_\mu^2 $, and
\begin{equation}
P^\nu (p, q) \equiv
\frac{m_\mu}{16 p^4 q^2}
(\pslash - \frac{1}{2} \qslash +m_\mu)
((\gamma^\nu \qslash - \qslash \gamma^\nu ) p^2 - 3 q^2 p^\nu )
(\pslash + \frac{1}{2} \qslash +m_\mu) .
\label{magproj}
\end{equation}
In the limit $q=0$ the $q^2$ term can be dropped in the denominator $V$
of (\ref{vertexdgrm}).
Then $V$ becomes a function of $p^2$ only and can be simplified to
\begin{equation}
V = \sum_{all~leptons} z_i m_i^2 -G ,
\end{equation}
where $G$ is defined by
\begin{equation}
G = -\frac{1}{2} p^\nu \left (\frac{\partial V}{\partial p^\nu} \right )_{q^2 =0,p^2=m_\mu^2}.
\end{equation}
$G$ can be reduced further to the form
\begin{equation}
G = \sum_{muon~only} z_i A_i m_\mu^2 ,
\end{equation}
by letting the external momentum $p$
flows through consecutive muon lines only.
This form is independent of how virtual photons are
attached to muon lines.
The information on photon attachment is contained in $A_i$.
This provides a significant simplification in programming.
Eq. (\ref{magmoment}) is the starting point of 
{\it Version B}.

In {\it Version A} 
the $g-2$ term is projected out from the RHS of (\ref{wtid2}).
In terms of Feynman parameters $z_1, z_2, \ldots z_N,~ (N=3n-1)$,
introduced in the self-energy-like diagram $G$, 
the $2n$-th order magnetic moment can be written as
\begin{eqnarray}
M^{(2n)} &=& \Biglb ( \frac{ -1 }{ 4 } \Bigrb )^n (n-1)! \int (dz)
\Biglb ( \frac{ \bm{E} + \bm{C} }{ n-1 } \frac{ 1 }{ U^2 V^{n-1} }
+ (\bm{N} +\bm{Z} ) \frac{ 1 }{ U^2 V^n } \Bigrb )~,
\label{WTtransform}
\end{eqnarray}
$\bm{E}, \bm{C}, \bm{N}$, and $\bm{Z}$ are pieces of the magnetic projection defined by
\begin{eqnarray}
{\bf N} &=& \frac{1}{4} Tr [P_1^\nu p_\nu (2G {\bf F})],  
~~~~ {\bf E} = \frac{1}{4} Tr [P_1^\nu {\bf E}_\nu ],   \nonumber  \\
{\bf C} &=& \frac{1}{4} Tr [P_2^{\mu \nu} {\bf C}_{\mu \nu} ], 
~~~~~~~~~ {\bf Z} = \frac{1}{4} Tr [P_2^{\mu \nu} {\bf Z}_{\mu \nu} ].  
\label{NECZ}
\end{eqnarray}
The factors $P_1^\nu$ and $P_2^{\mu \nu}$, derived from
the magnetic projector $P^\nu$ of (\ref{magproj}) by averaging over
the directions of $q_\mu$ subject to the constraint $q^\mu p_\mu = 0$,
are of the form
\begin{eqnarray}
P_1^\nu &=& \frac{1}{3} \gamma^\nu - \left(1 + \frac{4}{3} \frac{\pslash}{m_\mu}
\right) \frac{p^\nu}{m_\mu},  \nonumber   \\
P_2^{\mu \nu} &=& \frac{1}{3} \left(\frac{\pslash}{m_\mu} + 1\right)
\left(g^{\mu \nu} - \gamma^\mu \gamma^\nu + \frac{p^\mu}{m_\mu} \gamma^\nu 
- \frac{p^\nu}{m_\mu} \gamma^\mu \right).
\label{P1P2}
\end{eqnarray}

The operator ${\bf F}$ is the numerator factor of the self-energy-like diagram $G$:
\begin{equation}
{\bf F}= \gamma^{\alpha_1}
(\rlap{\hspace{0.07cm}/}{D}_1+m_1)
\gamma^{\alpha_2}
(\rlap{\hspace{0.07cm}/}{D}_2+m_2)
\ldots    
\gamma^{\alpha_{i}}
(\rlap{\hspace{0.07cm}/}{D}_{(2n-1)}+m_{(2n-1)})
\gamma^{\alpha_{j}}.
\label{numfactorF}
\end{equation}
Here $i$ and $j$ refer to the internal photon lines arriving at the
$(2n-1)-th$ and $2n$-th vertices along the lepton line
(which depend on the diagram $G$), ${\bf E}^\nu$ is defined by
\begin{equation}
{\bf E}^\nu \equiv \frac{\partial {\bf F}}{\partial p_\nu} =
\sum_{all~leptons} A_i {\bf F}_i^\nu ,
\label{numfactorE}
\end{equation}
and ${\bf F}_i^\nu$ is obtained from ${\bf F}$ by the substitution 
in the $i$-th line:
\begin{equation}
(\rlap{\hspace{0.07cm}/}{D}_i+m_i) \Longrightarrow \gamma^\nu .
\end{equation}
${\bf Z}_{\mu \nu}$ is defined by
\begin{equation}
{\bf Z}^{\mu \nu} = \sum_{j=1}^{2n-1} z_j {\bf Z}_j^{\mu \nu},
\label{numfactorZ}
\end{equation}
where ${\bf Z}_j^{\mu \nu}$ is obtained from ${\bf F}$ by the substitution
\begin{equation}
(\rlap{\hspace{0.07cm}/}{D}_j+m_j) \Longrightarrow 
\frac{1}{2} [\gamma^\mu \gamma^\nu (\Dslash_j +m_j )
- (\Dslash_j +m_j ) \gamma^\nu \gamma^\mu ]. 
\end{equation}
${\bf C}^{\mu \nu}$ is defined by
\begin{equation}
{\bf C}^{\mu \nu} = \sum_{i < j} C_{ij}  {\bf F}_{ij}^{\mu \nu},
\end{equation}
where
\begin{equation}
C_{ij} = \frac{1}{U^2} \sum_{k=1}^{2n-2}  \sum_{l=k+1}^{2n-1}
 z_k z_l (B_{ik}^{'} B_{jl}^{'}  
- B_{il}^{'} B_{jk}^{'}) ,
\end{equation}
and ${\bf F}_{kl}^{\mu \nu}$ is obtained from ${\bf F}$ by the substitution
\begin{equation}
(\Dslash_k + m_k ,~~~ \Dslash_l + m_l) \Longrightarrow (\gamma^\mu,~~~\gamma^\nu) 
\end{equation}
in the $k$-th and $l$-th lepton lines.
See (\ref{def0}) for the definition of $B_{ik}^{'}$.

Note that the potentially troublesome $q^2$ 
in the denominator of (\ref{magproj})
is absent in the formula (\ref{NECZ}) so that the limit
$q = 0$ can be taken without complication.
As a consequence (\ref{NECZ}) depends only on one external momentum $p$,
and the only scalar current needed are $A_i$ of the muon lines
associated with $p$.
When $A_i$ are expressed in terms of $B_{ij}$'s,
they have the same form for all diagrams irrespective of
how virtual photons are attached to the muon line.
This provides a useful simplification in programming.

We can now construct the integrand as follows:

(I) Express the integrand as a function of symbols
$B_{ij}$, $A_i$, $U$, $V$ for {\it Version B}
and additional $C_{ij}$ for {\it Version A}.
This can be achieved by an algebraic program 
(such as FORM \cite{form}) by 
which momentum integration is carried out exactly.
All integrals are generated from a small number of {\it templates}
by permutation of tensor indices of photon propagators.

(II)  Quantities $A_i$, $B_{ij}$, $C_{ij}$, etc., 
introduced in (I) are just symbols.
The next step is to turn them into explicit functions
of Feynman parameters. 
This is facilitated by a common template which generates
$B_{ij}$ and $U$ for all diagrams sharing the same topological structure.
Scalar currents $A_i$ (and $B_{ij}$)
must satisfy up to eight {\it junction laws}
and four {\it loop laws} for each diagram.
This provides very strong constraints on scalar currents
and sets up a powerful defense against trivial programming error.

\subsection{Construction of subtraction terms}
\label{renorm}

Following the discussion (c) at the beginning of this Appendix 
let us now discuss more explicitly how to construct UV-divergence
subtraction terms starting from the mother integral.

Suppose $M_G$ is the magnetic moment contribution of 
a proper vertex part of $2n$-th order defined by 
(\ref{vertexdgrm}) and (\ref{magmoment}).
Carrying out the $D$-operations in ${\bf F}_\nu$, one finds
\begin{eqnarray}
M_G &=& - (\frac{-1}{4})^n (n-1)~!
\int (dz)_G \left( \frac{F_0}{U^2 V^{n}} + \frac{F_1}{U^3 V^{n-1}} \right.  \nonumber \\
 &+& \left. \frac{F_2}{U^4 V^{n-2}} +\cdots+ \frac{F_{m_G}}{U^{m_G +2} V^{n-m_G}}\right),
\label{m_g}
\end{eqnarray}
where $m_G$ is the maximum number of contractions of $D$ operators,
which is equal to $n$ for a vertex part.
In (\ref{m_g}) the Feynman cut-off of photon propagators is not shown explicitly
for simplicity.

Suppose $M_G$ has a UV divergence when all loop momenta of a 
subdiagram $S$ consisting of $N_S$ lines and $n_S$ closed loops
go to infinity.
In the parametric formulation, this corresponds to the vanishing of 
the denominator $U$ when all $z_i \in S$ vanish simultaneously.
(Note that $z_i$ is a sort of conjugate to $k_i^2$
in a Laplace transform.)

To find how a UV divergence arises from $S$,
consider the part of the integration domain where $z_i ( \in S )$
satisfy $\sum_{i \in S} z_i \leq \epsilon$.
In the limit $\epsilon \rightarrow 0$, one finds
\begin{eqnarray}
V&=&{\cal O}(1),~~~~~~U = {\cal O}(\epsilon^{n_S}),  \nonumber  \\
B_{ij}&=&{\cal O}(\epsilon^{n_S -1})~~ {\rm if} ~~ i,j \in S,~  \nonumber  \\
B_{ij}&=&{\cal O}(\epsilon^{n_S })~~ {\rm otherwise}. 
\label{Slimit}
\end{eqnarray}
Let $m_S$ be the maximum number of contractions of $D$ operators
within $S$.
Simple power counting shows that the $(m+1)$-st term of $M_G$,
whose numerator has at most $m_S$ factors of $B_{ij}$, $i,j \in S$,
is divergent in the limit $\epsilon \rightarrow 0$ if and only if
\begin{equation}
N_S - 2n_S  \leq {\rm min}[ m, m_S ],
\label{div_crit}
\end{equation}
where ${\rm min}[m, m_S]$ means the lesser of $m$ and $m_S$.

If $S$ is a vertex part, we have $N_S = 3n_S$ and $m_S = n_S$.
If $S$ is an lepton self-energy part, we have $N_S = 3n_S -1$ and $m_S = n_S -1$.  
In both cases, (\ref{div_crit}) is satisfied only for
${\rm min} [m, m_S ] = m_S$, namely $m_S \leq m$.
This means that the UV divergence from $S$ is restricted to the
terms with $m_S$ contractions within $S$ in the last $m_G -m_S +1$
terms of (\ref{m_g}).

Let us now introduce a ${\bf K}_S$ operation, which extracts
the UV-divergent part (due to the subdiagram $S$)
of the Feynman integral 
\begin{equation}
M_G \equiv \int (dz)_G J_G
\end{equation}
in an analytically factorizable manner.
This is achieved by the following steps:

(a)  In the limit (\ref{Slimit}) keep only terms with the lowest power of 
$\epsilon$ in $U$, $B_{ij}$, $A_i$, $V$.
[$U$ then factorizes into a product of $U_S$ and $U_R$,
where $U_S$, $U_R$ are U-functions defined  on $S$, $R$, respectively,
and $R \equiv G/S$ is obtained from $G$ by shrinking $S$ to a point.
Similarly for $B_{ij}$.  $V$ is reduced to $V_R$.
Factorization of $A_i$ is more subtle since it has $U$
in its denominator.  The recipe here is to keep only those 
terms of $A_i$ whose numerator is a linear combination of
$B_{ij}$ with $i,j \in S$.
This is necessary for analytic factorization to work.]

(b)  Replace $V_R$ obtained in (a) by $V_R + V_S$, where $V_S$
is a $V$ function defined on $S$.
[Since $V_S = {\cal O} (\epsilon)$ whereas $V_R = {\cal O} (1)$,
this does not affect the leading singularity of the integrand in
the $\epsilon \rightarrow 0$ limit.

(c)  Rewrite $J_G$ in terms of the redefined parametric functions,
drop all terms except those with $m_S$ contractions within $S$,
and call the result ${\bf K}_S M_G$.

Since we deal in practice with logarithmic divergence only,
the steps (a), (b) and (c) are sufficient to ensure that
$(1- {\bf K}_S )M_G$ is convergent for $\epsilon \rightarrow 0$.
The inclusion of $V_S$ in (b) serves two purposes.  
One is to avoid spurious singularity which $V_R$ alone might
develop in other parts of the integration domain, and the other
is to enables us to decompose ${\bf K}_S M_G$ 
(assuming $S$ is a vertex part)
into  a product of lower order factors {\it analytically} as
\begin{equation}
{\bf K}_S M_G = {\hat L}_S M_{G/S},
\label{vertexrenorm}
\end{equation}
where ${\hat L}_S$ is the UV-divergent part of the renormalization
constant $L_S$.

If $S$ is a lepton self-energy part inserted between consecutive lepton lines $i$ and $j$
of $G$, we obtain a slightly more complicated result
\begin{equation}
{\bf K}_S M_G = \delta {\hat m}_S M_{G/S(i*)} +{\hat B}_S M_{G/S(i')}
\label{selfenergysubtr}
\end{equation}
Here ${\hat B}_S$ and $\delta {\hat m}_S$
are UV-divergent parts of renormalization constants
$B_S$ and $\delta m_S$.
Since ${\hat L}_S , M_{G/S} $, etc. 
are quantities of lower-orders,
they are already known or can be easily constructed from scratch.

Note that Eqs. (\ref{vertexrenorm})
and (\ref{selfenergysubtr}) are quite nontrivial since the LHS
are defined over the entire $n$-dimensional surface
while the RHS are products of integrals over lower-dimensional
spaces.
Identification of the second term on the RHS of (\ref{selfenergysubtr})
requires further work involving an integration by part.
(See \cite{kino0} for details.)

An IR divergence, which is caused by vanishing virtual photon momentum,
arises from the part of integration domain $R$ where $z_i$'s assigned to
the photon takes the largest possible value under the
constraint $\sum z_i = 1$.
All other $z_i$'s are pushed to zero in the IR limit.
Furthermore, the IR singularity, in order that it actually 
becomes divergent, must be enhanced by vanishing denominators of muon 
propagators  adjacent to the infrared photon.
In parametric language this corresponds to the vanishing of $V$
as $V \sim \delta^2$ for $ \delta \rightarrow 0$  in the integration domain characterized by

\[ z_i = \left\{  \begin{array}{ll}
                   O(\delta)   &  \mbox{if $i$ is a muon line in $R$} \\ 
                   O(1)        &  \mbox{if $i$ is a photon line in $R$} \\ 
                  O(\delta^2)  &  \mbox{if $i$ is in $G/R$.} 
                  \end{array}
\label{irlimit}
         \right.  \]

\vspace{4mm}

Starting from this one can obtain a power counting rule 
which enables us to identify
IR divergent terms in a way analogous to that of UV divergence.
The criteria to be satisfied by the IR subtraction operator ${\bf I}_R$ are

(i) it subtracts the IR divergent part of the mother integrand completely,

(ii) it factorizes analytically into a product of IR-divergent part
of renormalization constant and magnetic moment of lower orders.

The difference with ${\bf K}_S$ operation 
is that we must now look for
largest negative powers of $V$ instead of $U$ in (\ref{m_g}).
See \cite{kino0} for details.

\subsection{Diagrams containing a second-order lepton self-energy part}
\label{subsec:analyticitycheck}

When an integrand 
containing a second-order electron self-energy
part $S$ is expressed as a function of scalar currents,
all of its terms contribute to the UV divergence.
This means that the integrand of 
the  self-energy subtraction term,
{\it when expressed in terms of its own scalar currents}, 
has a form identical with that of mother integrand. 
Their difference comes solely from different structures of
scalar currents for mother and daughter integrals.
This provides the simplest example of (\ref{selfenergysubtr}).

To demonstrate it explicitly
let us go back to the vertex (\ref{vertexdgrm})
and rewrite ${\bf F}_\nu$ of (\ref{numfactor})
to exhibit the self-energy part $S$ explicitly:
\begin{eqnarray}
{\bf F}_\nu &=&
\gamma^{\alpha_1}
(\rlap{\hspace{0.07cm}/}{D}_1+m_1)
\gamma^{\alpha_2}
(\rlap{\hspace{0.07cm}/}{D}_2+m_2)
\ldots                                         
\gamma^{\alpha_{(i-1)}}
(\rlap{\hspace{0.07cm}/}{D}_{(i-1)}+m_{(i-1)})  \nonumber    \\
~~~~~~&~& \hspace*{2cm} \times 
\gamma^{\beta}
(\rlap{\hspace{0.07cm}/}{D}_{(i)}+m_{(i)})
\gamma_{\beta}     \nonumber   \\
& \times &
(\rlap{\hspace{0.07cm}/}{D}_{(i+1)}+m_{(i+1)})    
\ldots
\gamma^{\alpha_{(2n-1)}}
(\rlap{\hspace{0.07cm}/}{D}_{(2n)}+m_{(2n)})
\gamma^{\alpha_2n} ,
\label{numerator}
\end{eqnarray}
where
$\gamma^{\beta}
(\rlap{\hspace{0.07cm}/}{D}_{(i)}+m_{(i)})
\gamma_{\beta}$
comes from the second-order lepton self-energy part $S$.
($\gamma_{\nu}$ is not shown explicitly.)
Carrying out the contraction of $\gamma^{\beta}$ and $\gamma_\beta$
this factor can be written as 
$2m_{(i)}-2(\rlap{\hspace{0.07cm}/}{D}_{(i)}-m_{(i)})$. 
This leads naturally to the decomposition
\begin{equation}
{\bf F}_\nu =
{\bf F}_\nu^{(1)} +
{\bf F}_\nu^{(2)} ,
\label{selfmass}
\end{equation}
where ${\bf F}_\nu^{(1)}$ and ${\bf F}_\nu^{(2)}$ are obtained
by replacing $\gamma^\beta(\Dslash_{(i)}+m_{(i)})\gamma_\beta$ of (\ref{numerator}) 
$2m_{(i)}$ and $-2(\rlap{\hspace{0.07cm}/}{D}_{(i)}-m_{(i)})$,
respectively. 

As is easily seen the ${\bf F}_\nu^{(1)}$ part of the integral 
${\bf K}_S M_G$ 
factorizes exactly into a product of the self-mass $\delta m_2$
and the term $M_{G/S(i^* )}$,
which is obtained from $M_G$ by shrinking the lepton self-energy
diagram $S$ to a point:
\begin{equation}
\delta m_2 M_{G/S(i^* )}.
\end{equation}
In the same limit the 
${\bf F}_\nu^{(2)}$ part of the integral ${\bf K}_S M_G$ 
factorizes exactly into a
product of $\hat{B}_2$ (the UV-divergent part of $B_2$ (see \cite{kino0}))
and the term $M_{G/S(i^{'})}$,
which is obtained by inserting the factor
$-(\rlap{\hspace{0.07cm}/}{D}_{(i)}-m_{(i)})$ 
in $M_{G/S(i^* )}$:
\begin{equation}
\hat{B}_2 M_{G/S(i^{'})}.
\label{B2part}
\end{equation}
Note that in the ${\bf K}_S M_G$ operation leading to (\ref{B2part})
the contraction of $D_{(i)}$ with other $D$'s
in $G$ are dropped.

Finally, using an identity 
\begin{equation}
\int (dz)_G \frac{F_0}{U^2V^{N-2n}}
=- \int (dz )_G z_j \frac{\partial}{\partial z_j}
\left ( \frac{F_0}{U^2 V^{N-2n}} \right )  ,
\label{eq.5}
\end{equation}
which is a particular case of Eq. (A.7) of Nakanishi's Appendix
\cite{nakanishi},
one can transform
$M_{G/S(i^{'} )}$
into an amplitude $M_{G/[S,i+1]}$,
which is obtained by removing the self-energy part $S$ and
the adjacent lepton line $i+1$ from $G$.
(This is identical with a vertex of lower order
obtained directly by Feynman-Dyson rules.)

\subsection{An Illustration:
 Two-Step Renormalization of Fourth-Order Magnetic Moment}
\label{M_4}

\begin{figure}[ht]
\resizebox{9.5cm}{!}{\includegraphics{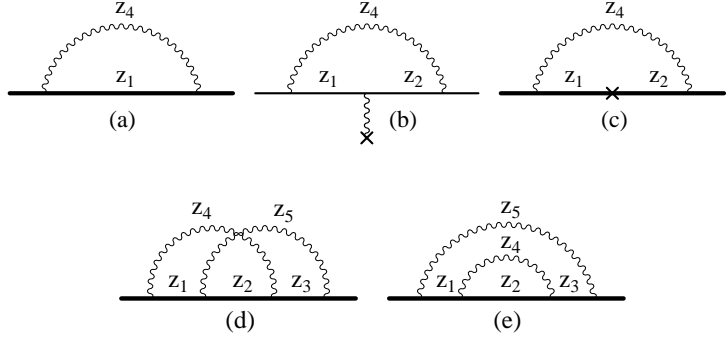}}
\vspace{0.5cm}
\caption{\label{fourth} 
Assignment of Feynman parameters $z_1 \ldots z_5$ is shown for
(a) second-order self-energy diagram,
(b) second-order vertex diagram,
(c) second-order self-energy diagram in which a 2-point vertex is inserted,
(d) fourth-order diagram $M_{4a}$,
(e) fourth-order diagrams $M_{4b}$.
Horizontal lines (except in (b)) are lepton lines in the presence
of the magnetic field.  Each of diagrams (d) and (e) represents
the sum of three fourth-order vertex diagrams.
}
\end{figure}

The formulation described above is illustrated here
by applying it to the fourth-order magnetic moment $M_4$
in {\it Version A},
which consists of two parts $M_{4a}$ and $M_{4b}$.
Eq. (\ref{WTtransform}) applied to
the diagram $G=[1,2,3,4,5]$
of Fig. \ref{fourth}(d) leads to
\begin{equation}
M_{4a} = \frac{1}{16} \int (dz) \left (  \frac{E_0 +C_0}{U^2 V}
+\frac{N_0 + Z_0}{U^2 V^2} +\frac{N_1 + Z_1}{U^3 V} \right ),
\label{eq.m4a}
\end{equation}
where , for simplicity, Feynman cut-off of photon lines is not shown
explicitly \cite{ck1,kino0}.
Numerator functions are expressed
in terms of scalar currents $A_i$ and $B_{ij}$:
\begin{eqnarray}
E_0 &=& 8 (2A_1 A_2 A_3 -A_1 A_2 - A_1 A_3 -A_2 A_3), ~~~ C_0 = -24 z_4 z_5/U ,  \nonumber  \\
N_0 &=& G(E_0 -8(2A_2 -1)),   \nonumber  \\
Z_0 &=& 8z_1 (-A_1+A_2+A_3+A_1A_2+A_1A_3-A_2A_3)   \nonumber   \\
    &+& 8z_2 (1-A_1A_2+A_1A_3-A_2A_3+2A_1A_2A_3)   \nonumber   \\
    &+& 8z_3 (A_1+A_2-A_3-A_1A_2+A_1A_3+A_2A_3),   \nonumber   \\
N_1 &=& 8G[B_{12} (2-A_3)+B_{13}(2-4A_2)+B_{23}(2-A_1)],   \nonumber   \\
Z_1 &=& -8z_1 [B_{12}(1-A_3)+B_{13}+B_{23}A_1]   \nonumber   \\
    &+& 8z_2 [B_{12}(1-A_3)-4B_{13}A_2+B_{23}(1-A_1)]   \nonumber   \\
    &-& 8z_3 [B_{12}A_3+B_{13}+B_{23}(1-A_1)], 
\label{enzc_a}
\end{eqnarray}
in addition to $z_1, z_2, z_3$, and $G$, and 
\begin{eqnarray}
B_{11} &=& z_{235},~~ B_{12} = z_{35},~~ B_{13} = -z_2,~~ B_{23}=z_{14}, \nonumber \\
B_{22} &=& z_{1345},~~B_{33}=z_{124},~~ U = z_{2} B_{12} +z_{14} B_{11},~~\nonumber \\
A_i &=& 1-(z_1 B_{1i}+z_2B_{2i}+z_3B_{3i})/U,~~~i=1,2,3,\nonumber \\
G &=& z_1A_1+z_2A_2+z_3A_3,~~~V=z_{123}-G,   \nonumber  \\
(dz) &=& \delta (1-z_{12345}) dz_1 dz_2 dz_3 dz_4 dz_5,~~~z_{235}\equiv z_2+z_3+z_5~.
\label{def6} 
\end{eqnarray}

The integral for $M_{4b}$ of Fig. \ref{fourth}(e) has the same form as (\ref{eq.m4a})
but has different definitions of functions \cite{ck1,kino0}:
\begin{eqnarray}
E_0 &=& 8 A_1 [4(A_2 -A_1) - A_1 A_2],  ~~~ C_0 = -8A_2,  \nonumber  \\
N_0 &=& -8G[4(1-A_1+A_1^2) + A_2 (1-4A_1 + A_1^2 )].   \nonumber  \\
Z_0 &=& 8z_{13} [4A_1 - A_2 (1+A_1^2 )] + 8z_2 A_2 (1+A_1^2 ),  \nonumber   \\
N_1 &=& 8G[8 (B_{11} - B_{12} ) + 3A_1 B_{12}],   \nonumber   \\
Z_1 &=& 24 (z_{13} -z_2 ) A_1 B_{12},
\label{enzc_b}
\end{eqnarray}
where
\begin{eqnarray}
B_{11} &=& z_{24},~~~ B_{12} = z_4,~~~, B_{22} = z_{1345},~~~ U = z_{135} B_{11} +z_2 B_{12},~~~\nonumber  \\
A_1 &=& z_5 B_{11}/U,~~~ A_2 = z_5 B_{12}/U,~~~ G = z_{13} A_1 + z_2 A_2, ~~~V = z_{123} - G.
\label{def8} 
\end{eqnarray}

The standard on-shell renormalized amplitudes $a_{4a}$ 
and $a_{4b}$ are defined by
\begin{equation}
a_{4a}=  M_{4a} -2 L_2 M_2
\label{standardrenorm_a} 
\end{equation}
and
\begin{equation}
a_{4b}=  M_{4b} -\delta m_2 M_{2^*} - B_2 M_2.
\label{standardrenorm_b} 
\end{equation}
We carry out the renormalization in two steps, 
expressing all quantities 
as parametric integrals.  For instance,
the magnetic moment $M_2$ is written in the form
(putting $m_4 = 0$ in $V$)
\begin{equation}
M_2 = -\frac{1}{4} \int (dz) \frac{N_0 + Z_0}{U^2 V},~~N_0+Z_0=4G(A_1-1) ,
\label{def_m2}
\end{equation}
where 
\begin{eqnarray}
B_{11}=1,~U&=&z_{14}B_{11},~A_1 = z_4/U,~G=z_1A_1,~ V=z_1+m_4^2z_4 - G, \nonumber \\
(dz)&=&\delta(1-z_{14})dz_1 dz_4, ~~~z_{14}\equiv z_1 + z_4.
\label{def1}
\end{eqnarray}

Following the general discussion the parametric integrals of
$B_2$ and $\delta m_2$ are split
into UV-divergent parts
$\hat{B}_2$ and $\delta \hat{m}_2$
and UV-finite part $\tilde{B}_2$ and $\delta \tilde{m}_2$:
\begin{equation}
B_2 = \hat{B}_2 + \tilde{B}_2,~~~~
\delta m_2 = \delta \hat{m}_2 +\delta \tilde{m}_2,
\label{splitting_1}
\end{equation}
where
\begin{eqnarray}
\hat{B}_2 &=& \frac{1}{4} \int (dz) \int_{\lambda^2}^{\Lambda^2} z_4 dm_4^2
\frac{E_0}{U^2 V}, ~ ~~E_0=-2A_1, 
\nonumber  \\
\tilde{B}_2 &=& \frac{1}{4} \int (dz) \frac{N_0}{U^2V}, ~~~N_0=2G(4-2A_1), 
\nonumber  \\
\delta \hat{m}_2 &=& \frac{1}{4} \int (dz) \int_{\lambda^2}^{\Lambda^2} z_4 dm_4^2
\frac {F_0}{U^2V} , ~~F_0=2(2-A_1),~~~~~\delta \tilde{m}_2 = 0.
\label{eq.def2}
\end{eqnarray}
$A_1$, $U$, $V$, $(dz)$ are the same as in (\ref{enzc_a}) and
$\Lambda$ and $\lambda$ are UV and IR cut-offs for the virtual
photon mass $m_4$. 
The UV cutoff is removed from $\tilde{B}_2$ since it is not UV-divergent.

Similarly, we split
the vertex renormalization constant $L_2$ for Fig. \ref{fourth}(b) as
$\hat{L}_2 + \tilde{L}_2$ ,where
\begin{eqnarray}
\hat{L}_2 &=& -\frac{1}{4} \int (dz) \int_{\lambda^2}^{\Lambda^2} z_4 dm_4^2
 \frac{F_1}{U^3V}, ~~~~F_1 = -2B_{11}, \nonumber  \\
\tilde{L}_2 &=& -\frac{1}{4} \int (dz) 
 \frac{F_0}{U^2 V},~~~F_0 = -2(1-4A_1+A_1^2),
\label{def3}
\end{eqnarray}
where
\begin{eqnarray}
B_{11}=1,~U&=&z_{124} B_{11},~A_1 = z_4/U,~G=z_{12}A_1,~V=z_{12}+m_4^2z_4 - G, \nonumber  \\
(dz)&=&\delta(1-z_{124})dz_1dz_2dz_4.
\label{def4}
\end{eqnarray}
Finally we need $M_{2^*}$ (magnetic moment contribution of the diagrams
in Fig. \ref{fourth}(c)):
\begin{eqnarray}
M_{2^*} &=& \frac{2}{4} \int (dz) \left ( 
 \frac{N_0^* + Z_0^*}{U^2V^2} 
+ \frac{N_1^*}{U^3V} 
+ \frac{E_0^*}{U^2V}
\right ),   \nonumber   \\
\vspace{4mm}
&~&N_0^*+Z_0^*=-8GA_1 (A_1 -1),~N_1^*=16GB_{11},~ E_0^*=-8A_1^2,
\label{def5}
\end{eqnarray}
where functions are the same as in (\ref{def4}).

\vspace{5mm}
\noindent
{\it (a) Two-step renormalization of $M_{4a}$.}

The first step is to rewrite (\ref{standardrenorm_a}) as
\begin{equation}
a_{4a}=  \Delta M_{4a} + (( {\bf K}_{12}+{\bf K}_{23} ) M_{4a}- 2 L_2 M_2) .
\label{2steprenorm_a} 
\end{equation}
${\bf K}_{12}$ is a UV-divergence extraction operator
for the vertex subdiagram 
$S \equiv [1,2,4]$.  
The integral  
\begin{equation}
\Delta M_{4a}  = (1 - {\bf K}_{12}-{\bf K}_{23} ) M_{4a} 
\label{def7} 
\end{equation}
is UV-finite by construction.
${\bf K}_{12} M_{4a}$ can be written as
\begin{equation}
{\bf K}_{12} M_{4a} = \frac{1}{16} \int (dz) 
 \int_{\lambda^2}^{\Lambda^2} z_4 dm_4^2 
\frac{N^{'}_1 + Z^{'}_1}{U^{'3} V^{'2}} ,
\label{eq.k12_m4a}
\end{equation}
where the Feynman cut-off of photon mass $m_4$ is shown explicitly and
\begin{eqnarray*}
&&B_{12} = z_{35},~~U^{'}=z_{124}B_{12},~~A^{'}_3=z_5/z_{35},~~A_1^{'}=z_4/z_{124}
\nonumber \\
&&G^{'}=z_3A^{'}_3,~~V^{'}=z_{123} +m_4^2 z_4 -G^{'}-z_{12}A_1^{'},~~~\nonumber \\
&&N^{'}_1 + Z^{'}_1 = 8G^{'}B_{12}(1-A^{'}_3).
\end{eqnarray*}
${\bf K}_{23} M_{4a}$ for the vertex diagram [2,3,5] can be constructed in a similar way.

To show that (\ref{eq.k12_m4a}) can be factorized and reduced to the 
RHS of (\ref{vertexrenorm}),
let $T\equiv [3,5]$ be the reduced diagram obtained from $G$ by shrinking
$S$ to a point.
Let us define $z_S \equiv z_{124}, z_T \equiv z_{35}$, and
introduce Feynman parameters $x_i$ and $y_j$ for the sets $S$ and $T$
in such a way that
\begin{equation*}
x_1 = z_1/ z_S,~~ x_2 = z_2/ z_S,~~ x_4 = z_4/ z_S,
~~ y_3= z_3/z_T, ~~ y_5= z_5/z_T.
\end{equation*}
Then, after few steps of manipulation, 
which amounts to the substitution
\begin{equation*}
B_{12} \rightarrow  B_{12}^T (=1), ~~ A_3^{'}\rightarrow A_3^T (=y_3/y_{35}),~~G^{'} \rightarrow y_3 A_3^T,~~
\end{equation*}
\begin{equation*}
U^{'} \rightarrow U_S U_T,~~
V^{'} \rightarrow z_S V_S + z_T V_T,~~
N_1^{'} + Z_1^{'} \rightarrow 8 y_3 A_3^T (1-A_3^T),~~
\end{equation*}
with
\begin{equation*}
 A_1^S = x_4/x_{124}, ~~U_S = z_S,~~U_T = z_T,~~ 
\end{equation*}
\begin{equation*}
V_S = x_{12} + m_4^2 x_4 -G_S,~~ V_T = y_3-G_T,
~~~G_S=x_{12} A_1^S,~~ G_T =y_3 A_3^T,
\end{equation*}
we can rewrite (\ref{eq.k12_m4a}) as
\begin{equation}
{\bf K}_{12} M_{4a} = \frac{1}{16}\int (dx) \int (dy)
 \int (d\tilde{z}) \int_{\lambda^2}^{\Lambda^2} x_4 dm_4^2 
\frac{8 y_3 A_3^T (1-A_3^T)}{(z_S V_S + z_T V_T)^2} ,
\label{k12_m4a}
\end{equation}
where $(d\tilde{z}) \equiv \delta(1-z_S-z_T)dz_Sdz_T$.
Note that $U^{'}$ in (\ref{eq.k12_m4a})
factorized as $U_S U_T$,
and they canceled out completely against $z_S$ and $z_T$
in the numerator.
Since $z_S$ and $z_T$ now appear linearly in the denominator only,
the $z$-integration can be carried
out explicitly using Eq. (\ref{feynmanformula})
which reduces the integral (\ref{k12_m4a})
to a product of $\hat{L}_2$ and $M_2$:
\begin{equation}
{\bf K}_{12} M_{4a} = \hat{L}_2 M_2.
\label{factorized} 
\end{equation}
This is a particular case of (\ref{vertexrenorm}).
Similarly for ${\bf K}_{23} M_{4a}$.  
Note that ${\bf K}$-operation extracts only the UV-divergent part $\hat{L}_2$ 
of $L_2$.
This exact factorization of (\ref{factorized})
is generalizable to
all orders, which is crucial for carrying the
two-step renormalization scheme over to higher orders.

From 
(\ref{standardrenorm_a}), 
(\ref{def3}), and (\ref{factorized})
we obtain
\begin{equation}
a_{4a}= \Delta M_{4a} -2 \tilde{L}_2 M_2,
\label{int_renorm} 
\end{equation}
where both terms on the right-hand-side are UV-finite.
($\tilde{L}_2$ is IR-divergent.)

\vspace{5mm}
\noindent
{\it  (b) Two-step renormalization of $M_{4b}$.}

We begin by rewriting (\ref{standardrenorm_b}) as
\begin{equation}
a_{4b}=  \Delta^{'} M_{4b} + ( {\bf K}_2 M_{4b}- \delta m_2 M_{2^*} - B_2 M_2).
\label{2steprenorm_b} 
\end{equation}
${\bf K}_2$ isolates the UV divergence of $M_{4b}$ arising from the 
self-energy subdiagram $S \equiv [2,4]$. 
By construction the integral
\begin{equation}
\Delta^{'} M_{4b}  = (1 - {\bf K}_2 ) M_{4b} 
\label{definition5} 
\end{equation}
is UV-finite.
Following the rule given in Appendix \ref{renorm},
${\bf K}_2 M_{4b}$  is obtained 
by dropping terms in $Z_0$ and $Z_1$ of (\ref{enzc_b}) 
that contains $z_2$ explicitly
($z_2$ and $z_4$ hidden within $A_i$ and $B_{ij}$
must be kept to the leading order in the ${\bf K}_2$ limit.)
Noting that $A_2 \rightarrow A_1^{'} A_2^{'}$ in the ${\bf K}_2$-limit
we find from (\ref{enzc_b})
\begin{eqnarray}
E^{'}_0 &=& 8 A^{'2}_1 [4(A^{'}_2 -1) - A^{'}_1 A^{'}_2],  
~~~ C^{'}_0 = -8A^{'}_1 A^{'}_2,  \nonumber  \\
N^{'}_0 &=& -8G^{'} [4(1-A^{'}_1+A^{'2}_1) +A^{'}_1 A^{'}_2 (1-4A^{'}_1 + A^{'2}_1)].   \nonumber  \\
Z^{'}_0 &=& 8z_{13} A^{'}_1 [4 - A^{'}_2 (1+A^{'2}_1 )] ,  \nonumber   \\
N^{'}_1 &=& 8G^{'} [8 (B^{'}_{11} - B^{'}_{12} ) + 3A^{'}_1 B^{'}_{12}],   \nonumber   \\
Z^{'}_1 &=& 24 z_{13}  A^{'}_1 B^{'}_{12}, 
\label{enzc_Ks}
\end{eqnarray}
where
\begin{equation*}
A^{'}_1 = z_5/z_{135},  ~~~ A^{'}_2 = z_4/z_{24}, ~~~
  G^{'} = z_{13} A^{'}_1,
\end{equation*}
Including the regulator mass $m_4$ of photon 4
explicitly, we can express ${\bf K}_2 M_{4b}$ as
\begin{equation}
{\bf K}_2 M_{4b} = \frac{1}{16} 
    \int (dz) z_4 \int_{\lambda^2}^{\Lambda^2}  dm_4^2 
\left [ \frac{E^{'}_0 + C^{'}_0}{U^{'2} V^{'2}}
+ \frac{2(N^{'}_0 + Z^{'}_0)}{U^{'2} V^{'3}} 
+ \frac{N^{'}_1 + Z^{'}_1}{U^{'3} V^{'2}} \right ].
\label{eq.ksm4b}
\end{equation}
In the ${\bf K}_2$-limit $U^{'}$  and $V^{'}$ decompose as
\begin{equation}
U^{'} \rightarrow  U_S^{'} U_T^{'},
~~~V^{'} \rightarrow  z_S V_S^{''} + z_T V_T^{''},
\label{UsUt}
\end{equation}
where $ U_S^{'} = z_S =z_{24}, U_T^{'} = z_T = z_{135}$,
and $V_S^{''}$ and $V_T^{''}$ are $V$ functions for the subdiagrams
$S \equiv [2,4]$ and $T \equiv [1,3,5]$
to be made explicit in the following.

Let us introduce Feynman parameters $x_i$ and $y_i$ 
for the subdiagrams $S$ and $T$ as follows:
\begin{equation*}
x_2 = z_2/z_S,~~~x_4 = z_4/z_S,~~~y_1 = z_1/z_T,~~~y_3 = z_3/z_T,
~~~y_5 = z_5/z_T,
\end{equation*}
and rewrite (\ref{eq.ksm4b}) in terms of 
\begin{equation*}
B_{11}^{'} = x_{24} z_S,~~ B_{12}^{'} = x_4 z_S,~~
A_1^{'} = y_5 /y_{135},~~ A_2^{'} = x_4/x_{24},
~~ G_S = x_2 A_2^{'},~~G_T = y_{13} A_1^{'}, ~~
\end{equation*}
\begin{equation*}
U_S^{'} = z_S U_S^{''}, ~~~U_S^{''}  =x_{24},
~~~U_T^{'} = z_T U_T^{''},~~~U_T^{''}  = y_{135} ,
\end{equation*}
\begin{equation*}
V^{'} = z_S V_S^{''} +z_T V_T^{''},
 ~~~V_S^{''}  =x_2+m_4^2 x_4 -G_S,~~~V_T^{''}  = y_{13}- G_T,
\end{equation*}
\begin{equation}
(dx) = \delta (1-x_{24}) dx_2 dx_4 ,
~~~(dy) = \delta (1-y_{135})dy_1 dy_3 dy_5.
\label{y-k2m4b}
\end{equation}
Then, dropping superscripts $'$ and $''$ for simplicity, we can re-express (\ref{eq.ksm4b}) as
\begin{eqnarray}
{\bf K}_2 M_{4b} &=& \frac{1}{16} \int (dx) \int (dy)
    \int (d\tilde{z}) \int_{\lambda^2}^{\Lambda^2} x_4  dm_4^2 \nonumber  \\
&[& \frac{z_S^{2-2} z_T^{2-2}}{U_S^2 U_T^2}  
 \frac{8 A_1^2 (4 (A_2-1)-A_1 A_2 )-8 A_1 A_2}{(z_S V_S + z_T V_T )^2} 
\nonumber \\
&+& \frac{z_S^{2-2} z_T^{3-2}}{U_S^2 U_T^2}  
 \frac{-16 G_T  (4 (1-A_1 +A_1^2)+ A_1 A_2 (1-4A_1+A_1^2)
- (4-A_2 (1+A_1^2)))}{(z_S V_S + z_T V_T )^3}   \nonumber \\
&+& \frac{z_S^{3-3} z_T^{3-3}}{U_S^3 U_T^3} 
\frac{8 G_T (8 (B_{11} - B_{12} ) + 3 A_1 B_{12}+3B_{12})}{(z_S V_S + z_T V_T )^2} ],
\label{eq.ksm4b2}
\end{eqnarray}
where
\begin{equation*}
(d \tilde{z}) = \delta (1-z_S - z_T) dz_S dz_T.
\end{equation*}

Now the $\tilde{z}$ integration can be carried out exactly 
using (\ref{feynmanformula}). The result is 
\begin{eqnarray}
{\bf K}_2 M_{4b} &=& \frac{1}{16} \int (dx) \int (dy)
    \int_{\lambda^2}^{\Lambda^2} x_4 dm_4^2
 [ \frac{ 8 A_1^2 (4 (A_2 -1)-A_1 A_2) -8A_1 A_2}{U_S^2 V_S U_T^2 V_T } 
\nonumber  \\
&+&\frac{-8G_T (4 (1-A_1+A_1^2 )+ A_1 A_2 (1-4 A_1 +A_1^2)
- (4-A_2 (1+A_1^2)))}{U_S^2 V_S U_T^2 V_T^2}   
\nonumber  \\
&+& \frac{8 G_T B_{11} (8 (1-A_2 ) +3 A_1 A_2 + 3 A_2)}
{U_S^3  V_S U_T^3 V_T } ],
\label{eq.ksm4b3}
\end{eqnarray}
where $B_{12}/B_{11} = A_2$ is used.
Decomposing the numerator of each term 
into terms proportional to $(2-A_2)$ and $A_2$
following (\ref{selfmass}),
we can rewrite (\ref{eq.ksm4b3}) as
\begin{eqnarray}
{\bf K}_2 M_{4b} &=& \frac{1}{16} \int (dx) 
    \int_{\lambda^2}^{\Lambda^2} x_4 dm_4^2 
     \frac{2-A_2}{U_S^2 V_S} \nonumber  \\
& &\times \int (dy)
 \left[ \frac{ -16 A_1^2 }{U_T^2 V_T } 
+ \frac{-16 G_T (-A_1+A_1^2 )}{U_T^2 V_T^2} 
+ \frac{32 G_T B_{11}}{U_T^3 V_T } \right]  
\nonumber   \\
&+& \frac{1}{16} \int (dx)
    \int_{\lambda^2}^{\Lambda^2} x_4 dm_4^2 
\frac{A_2}{U_S^2 V_S} 
\nonumber  \\
& & \times \int (dy)
[ \frac{ 8 A_1^2 (2-A_1)- 8 A_1}{U_T^2 V_T }   
+\frac{-8G_T (1-A_1-A_1^2+ A_1^3)}{U_T^2 V_T^2} \nonumber  \\
&+& \frac{2 G_T B_{11} (-4 +12 A_1) }{U_T^3 V_T } ].
\label{eq.ksm4b4}
\end{eqnarray}
Note that both terms are now just products of $x$-integral and $y$-integral.
Comparing them with (\ref{eq.def2}) and (\ref{def5}), it
is easy to see that the first term is 
$\delta m_2 M_{2*}$, which corresponds 
to the first term of (\ref{selfenergysubtr}).
The $x$-integral of the second term is proportional to $\hat{B}_2$.
Thus the remaining task is to identify the $y$-integral
with $M_2$ in (\ref{eq.def2}).
The first step of demonstration is to transform $M_2$ by
an integration-by-part.
Dropping the suffix $T$ for simplicity, we obtain
\begin{eqnarray}
M_2 &=& \int (dy) \frac{y_1 A_1 (1-A_1 )}{U^2 V} \nonumber  \\
    &=& -\int (dy)  y_1 \frac{\partial}{\partial y_1} \left ( \frac{y_1 A_1 (1-A_1 )}{U^2 V} \right ) \nonumber  \\
    &=& - \int (dy) y_1 \left[\frac{A_1 (1 - A_1)}{U^2 V}
      +\frac{y_1 B_{11}(-3 A_1 + 4 A_1^2)}{U^3 V}
      - \frac{y_1 A_1 (1-A_1 )(1-A_1^2)}{U^2 V^2}\right],
\label{m2trans1}
\end{eqnarray}
where we have used
\begin{equation*}
B_{11} =1,~~~U=y_{14} B_{11},~~~A_1 = \frac{y_4 B_{11}}{U},
~~~\frac{\partial U}{\partial y_1} = B_{11},
~~~\frac{\partial A_1}{\partial y_1} = - \frac{A_1 B_{11}}{U},
~~~\frac{\partial V}{\partial y_1} = 1-A_1^2.
\end{equation*}
Noting that $y_1 B_{11}/U = 1 -A_1$,
(\ref{m2trans1}) can be reduced to 
\begin{equation}
M_2 =  \int (dy) y_1 \left[\frac{2 A_1 (1 - A_1)(1-2A_1)}{U^2 V}
      + \frac{y_1 A_1 (1-A_1 )(1-A_1^2)}{U^2 V^2}\right].
\label{altform2}
\end{equation}
It is now easy to see that (\ref{altform2})
is proportional to the second $y$-integral
of ({\ref{eq.ksm4b4}). 
One has only to note that
$(dy)y_1 = \delta(1-y_1-y_4)y_1 dy_1 dy_4 $ in (\ref{altform2}) 
is equivalent to $(dy) = \delta(1-y_1-y_3-y_5)dy_1dy_3dy_5$
in (\ref{eq.ksm4b4})
since its integrand depends on the sum $y_1+y_3$ only.
Eq. (\ref{m2trans1}) is a special case of (\ref{eq.5}).

Making use of formulas for $\delta m_2$, $\hat{B}_2$,
and $M_{2^*}$ in (\ref{eq.def2}) and (\ref{def5}), we thus obtain
\begin{equation}
{\bf K}_2 M_{4b} =\delta m_2 M_{2^*} + \hat{B}_2 M_2,
\label{k2m4b} 
\end{equation}
which is a particular case of (\ref{selfenergysubtr}).

Using 
(\ref{2steprenorm_b}) 
and 
(\ref{definition5}) 
$a_{4b}$ in
(\ref{standardrenorm_b}) 
can be rewritten as
\begin{equation}
a_{4b}= \Delta^{'} M_{4b} - \tilde{B}_2 M_2.
\label{renorm_b} 
\end{equation}
Both terms on the right-hand-side are UV-finite
although $\Delta^{'} M_{4b}$ is IR-divergent.

\vspace{5mm}
\noindent
{\it  (c) Separation of IR divergence in $\Delta^{'} M_{4b}$.}

The sum $a_{4a}+a_{4b}$ is UV- and IR-finite.
However, for numerical evaluation it is necessary
to remove IR-divergent terms from 
the integral $\Delta^{'} M_{4b}$.
This is where Step 2 comes in.
A general procedure for isolating the IR divergent terms 
is given in \cite{kino0}.
Power-counting shows that,
of three vertex diagrams contributing to
$\Delta^{'} M_{4b}$, IR divergence arises only
from the vertex in which the magnetic field acts on the
muon line denoted $z_2$ in Fig. \ref{fourth}(e). 
More specifically, it arises from the domain
\begin{equation}
z_5 = 1 - {\cal O} (\delta), ~~~z_1, ~z_3 = {\cal O} (\delta), ~~~z_2,~z_4 = {\cal O}(\delta^2),
\end{equation}
where $\delta \sim \lambda$ and $\lambda$ is the photon mass.
In this domain one may define an IR-extracting operator
${\bf I}_T$ by
\begin{eqnarray}
U \rightarrow U_S U_T,~~V \rightarrow V_S+V_T,~~V_S&=&z_2 (1-A_2),~~V_T = z_{13} (1-A_1) + z_5 m_5^2 .
 \nonumber  \\
A_1 = z_5/z_{135},~~ A_2 = z_4/z_{24},~~U_S&=&z_{24},~~U_T=z_{135}. 
\label{IRlimit}
\end{eqnarray}
This  definition is actually identical with that of the ${\bf K}_2$-limit
(\ref{eq.ksm4b}) except that the substitution$ A_2   \rightarrow A_1^{'}
A_2^{'}$ (see above Eq. (\ref{enzc_Ks}) ) is replaced by $A_2  \rightarrow  A_2^{'} ( \Rightarrow A_2) $.

The separation of IR-divergent term of $\Delta^{'} M_{4b}$ may be written as
\begin{equation*}
\Delta^{'} M_{4b} = \Delta M_{4b} + {\bf I}_T \Delta^{'} M_{4b},
\end{equation*}
where
\begin{equation}
\Delta M_{4b}= (1-{\bf I}_T) \Delta^{'} M_{4b} .
\label{IRlimit2}
\end{equation}

The definition (\ref{IRlimit}),
although it picks up the IR-singularity correctly,
is actually not complete until an 
appropriate numerator function is chosen.
We chose to define the 
IR separation operator ${\bf I}_T$, $T\equiv [1,3,5]$, by
\begin{equation}
{\bf I}_T \Delta^{'} M_{4b} = \frac{1}{16} \int (dz) \frac{F_{T}F_S}{U^2V^2},
\label{IRterm}
\end{equation}
where
\begin{equation}
F_{T} =-2(1-4A_1+A_1^2),~~ F_S = -4 z_2 A_2 (1-A_2).
\end{equation}
$F_T$ is chosen to coincide with $F_0$ of (\ref{def3})
and $F_S$ corresponds to $N_0+Z_0$ of (\ref{def_m2}).
After taking steps analogous to (\ref{UsUt}) and (\ref{y-k2m4b}), 
this choice leads to an exact factorization of
the integral (\ref{IRterm}) as
a product of known second-order integrals
\begin{equation}
{\bf I}_T \Delta^{'} M_{4b} = \tilde{L}_2 M_2.
\end{equation}
Note that a particular choice of numerator is not crucial
as far as it deals with the IR divergence correctly
since what is subtracted must be put back in the end.

Summing up all terms we obtain
\begin{eqnarray}
a_4 &=& a_{4a}+a_{4b} \nonumber  \\
    &=& \Delta M_{4a} + \Delta M_{4b} -\Delta B_2 M_2
\label{a4}
\end{eqnarray}
where $\Delta B_2 \equiv \tilde{L}_2+\tilde{B}_2 (= 3/4)$
and  $M_2 = 1/2.$  Numerical evaluation of
$\Delta M_{4a}$ and $\Delta M_{4b}$ gives 
\begin{eqnarray}
\Delta M_{4a} &=& ~~0.218~342~(17),       
\nonumber  \\
\Delta M_{4b} &=& -0.187~501~(14),      
\nonumber  \\
a_4 &=& -0.344~158~(22),   
\end{eqnarray}
which is an update of  the old evaluation \cite{kino0}.
It is in good agreement with the analytical value -0.344~166 ... .

We described the fourth-order case in full detail
because it will serve as a good prototype for higher-order cases.
To begin with integrals such as {\ref{eq.m4a})
with integrands (\ref{enzc_a}) and (\ref{enzc_b})
are obtained by a simple algebraic program written in FORM.
Subsequent manipulation of integrands proceeds in a well-organized manner.
The important point is that 
all higher-order integrals can be handled in the same manner,
the necessary extension being straightforward.
This is the reason why we are able to treat the algebra of 
higher-order cases with complete confidence.

\section{Non-statistical error in numerical evaluation of Feynman Integral}
\label{ddproblem}

Our integrand  of Group IV is an algebraic function of more than 4,000 terms,
each term being a product of up to 10 functions defined
on a unit 10-dimensional cube:
\begin{equation}
0 \leq  x_i  \leq 1 , ~~~ i = 1, 2, \cdots , 10 .
\label{hypercube}
\end{equation}
FORTRAN codes of some integrals are as large as 100 kilobytes.
These integrals are identical with the 
corresponding integrals for the electron vertices, the only difference being
the value of the parameter $m_e /m_\mu$.
However, the behavior of muon integrals are strongly influenced
by the presence of a singularity located at a distance
of order $(m_e /m_\mu )^2$ just outside of the
integration domain (\ref{hypercube}), i. e., a unit cube.
This makes numerical integration 
of some integrals more delicate or difficult
compared with the electron case.
This is the main (though not the only) source 
of the $d$-$d$ problem in the muon $g-2$ calculation.


Numerical integration of these integrals is carried out 
using an adaptive-iterative Monte-Carlo integration routine
VEGAS \cite{lepage}. 
It is the only effective method currently available
to integrate such huge integrals.
It is an adaptive-iterative integration routine
based on random sampling of the integrand.
In the $i$-th iteration, the integral is evaluated by sampling
it at points chosen randomly according to a
distribution $\rho_{i-1}$ 
(a step function defined by grids) constructed in the $(i-1)$-st iteration.
This generates an approximate value $I_i$ of the integral,
its uncertainty $\sigma_i$, and the new distribution
function $\rho_i$ to be used in the next iteration.
The distribution $\rho_i$ is constructed in such a way that the grids
concentrate in the region where the integrand is large.
The construction of 
$\rho_i$ in the $(i - 1)$-st iteration 
involves a positive parameter $\beta$
that controls the speed of $``$convergence" to a stable configuration.
In most cases we chose $\beta = 0.5$.
We may even be forced to choose $\beta = 0$ (no change in $\rho$), 
which is necessary in some difficult cases.

After several iterations $I_i$ and $\sigma_i$ 
are combined assuming that all iterations 
are statistically independent.
The combined value and uncertainty are given by
\begin{equation}
I = (\sum_i ( I_i /\sigma_i^2 ))/(\sum_i (1/ \sigma_i^2 )), ~~~
\sigma = (\sum_i (1/ \sigma_i^2 )) ^{-1/2} .
\label{combines}
\end{equation}
For well-behaved integrals 
$\rho_i$ converges rapidly 
to a (practically) stable configuration.
Once $\rho_i$ is stabilized, the error generated by VEGAS
is (nearly) statistical and
proportional to ${\cal N}^{-1/2}$,
where ${\cal N}$ is the total number of data samplings.

After the point-by-point renormalization is made
the integrand has the form 
\begin{equation}
f = f_0 + \cdots + f_r  ,
\label{integrandf}
\end{equation}
where $f_0$ is obtained directly from a Feynman diagram
and $f_1$,..., $f_r$ are terms needed to renormalize
UV (and/or) IR divergences of $f_0$.
Terms $f_0$, ..., $f_r$ are all divergent 
on the surface of the unit cube (\ref{hypercube}).
The sum $f$ is mathematically well-defined and integrable.

This does not guarantee, however, that $f$ is well-behaved 
on a computer.
This is because expressions for $f_0$,..., $f_r$ on computer 
are only as accurate as the number of digits in use 
(64 bits, 128 bits, etc.).
In the part of the domain where
$f_0$, ..., $f_r$ are singular, $f$
loses most or all of significant digits and is
affected severely by round-off errors.
When this happens, $I_i$ and $\sigma_i$
become unreliable or even divergent.
Note that this problem is an {\it inevitable} consequence of 
any computer calculation in which only a finite number of
digits is available.
We shall call it the digit-deficiency or $d$-$d$ problem.
In order to cope with the $d$-$d$ problem before it 
upsets the integration, we have developed several strategies.

\vspace{2mm}
\noindent
{\it a. Stretching.}
The integrand $f$ defined in (\ref{integrandf}) may still have  
integrable singularities on some boundary surfaces,
which can be removed by an appropriate change of
variables.
However, 
it is difficult to find analytically correct mapping
because of the complicated structure of the integrand.
A simple way to remove or 
weaken the $d$-$d$ problem is the $``$stretching"
defined as follows:
Suppose VEGAS finds after several iterations that 
the integrand samplings tend to
concentrate in the vicinity of an $(n-1)$-dimensional surface, say 
$x_1 = 0$, perpendicular to the $x_1$ axis.
Then the mapping
\begin{equation}
x'_1 = x_1^{a_1} ,
\end{equation}
where $a_1$ is a real number greater than 1,
stretches out the
domain near $x_1 = 0$ and random samplings 
in the $x'_1$ variable give more attention to this region
from the beginning of iteration. 
Also, the Jacobian $a_1 x^{a_1 - 1}$ of this mapping
weakens the singularity.
Similarly, the singularity at $x_1 = 1$ can be weakened by the stretching
\begin{equation}
x'_1 = 1 - (1 - x_1)^{b_1} ,~~~ b_1 > 1.
\end{equation}
Stretching is a one-to-one mapping of a unit hypercube onto itself.
It may be applied to all variables independently.
An appropriate stretching speeds up
convergence of $\rho$ considerably.
Note also that different stretchings lead to statistically independent
samplings of an integral which must give the same
answer within error bars.
This flexibility is important in assessing the reliability
of results of integration.
Of course, stretching does not always work well since
it disregards the actual (and hard-to-identify)
analytic structure of the integrand.

\vspace{2mm}
\noindent
{\it b. Splitting.}
Going from double precision (real*8) to 
quadruple precision (real*16) (or even higher) arithmetic 
is the most effective way 
to control the $d$-$d$ problem.
One practical obstacle is that real*16 slows down computation
by a factor 20 $\sim$ 30. Thus we were not able to use  real*16
extensively until massively parallel computers became readily available.

Actually, in many cases, real*16 is needed only in a small part
of the integration domain.
It is therefore useful to adopt the following strategy:
Start the evaluation of a Feynman integral
in real*8, which explores the integrand at high speed. 
If it identifies the region causing the $d$-$d$ problem,
split the integration domain into a small (rectangular) part 
in which the $d$-$d$ problem occurs and the remainder.
The difficult region is then evaluated in real*16,
while the rest continues in real*8.
This strategy has been very successful and most integrals
have been evaluated in this manner.

Recently, a modified algorithm 
of VEGAS has been developed which 
makes this splitting local and automatic \cite{sinkovits}.
In this approach the integrand $f$ is first evaluated
at each point in real*8.
The result is tested by computing the ratio
\begin{equation}
t = (f_+ + | f_- |) /  | f_+ + f_- | ,
\label{tratio}
\end{equation}
where $f_+$($f_-$) is the sum of positive(negative) terms
of $f_0, ... f_r$.
If $t$ is larger than an empirically selected number $t_0$,
it signals a possible $d$-$d$ problem.
The integrand is then reevaluated in real*16 at the same spot.
If the $d$-$d$ problem is not severe, this method is very efficient and
runs much faster than pure real*16.
In difficult cases, however, a simple splitting may
work faster since it does not require the overhead
needed in computing (\ref{tratio}).

\vspace{2mm}
\noindent
{\it c. Freezing.}
Sometimes, it is very difficult to find a reasonable stretching that 
does not run into the $d$-$d$ problem
before it settles down to a (nearly) stable $\rho$.
In such a case, one may freeze $\rho$
by putting $\beta = 0$ few steps before the $d$-$d$
problem becomes serious.
The resulting $\rho$ is not optimal so that it requires
longer computing hours to achieve the desired
statistical uncertainty.

\vspace{2mm}
\noindent
{\it d. Chopping.}
If procedures $a$, $b$, $c$ fail to solve the $d$-$d$ problem,
one may restrict 
some integration axis (0, 1) to $(\delta , 1 - \delta )$,
where $0 \leq \delta \ll 1$, to
exclude the danger zone.  This is referred to as {\it chopping}.
The error introduced by $chopping$ is of order
$\delta^{1/2} (\ln \delta )^a$, where the positive number $a$ 
can not be fixed without knowing
the analytic structure.
In practice it is sufficient to find a crude value of $a$
by carrying out integration for several values of $\delta$.
When $chopping$ is used, we must 
carry out full scale calculations for several $\delta$
(which require extra computing time).
Note also that integration becomes more and more difficult as 
$\delta$ gets smaller.
The difficulty in assessing the effect of chopping was
the major source of non-statistical uncertainty
in earlier calculations.

Chopping can produce a crude approximate result vary rapidly.
Thus it was used in early stage of our work to obtain estimates of a rough 
order of magnitude.
However, it turned out to be not effective for obtaining
more precise results.
Thus it was abandoned entirely in the later phases of our work.

Our final results were obtained using $stretching$,
$splitting$, and $freezing$, or their combinations.
In most cases $stretching$ and $splitting$ are sufficient to solve the $d$-$d$
problem.
In some cases, however, even $splitting$ was not sufficient.
In the absence of higher precision arithmetic, the only
effective way to control the $d$-$d$ problem was 
$freezing$.
In such a case it is still useful to divide the integration
domain into several pieces and apply $freezing$ in only
one of them.
Moreover, it is found to be
useful to restrict the number of samplings per iteration
to a relatively small number and use a very large number of iterations.
This will enable us to accumulate large statistics while
controlling the amount of wasted iterations to an acceptable level.



\begin{thebibliography}{99}
\bibitem{dimensionalreg}
An alternative is dimensional regularization.
We do not consider it here since it is not known how to
implement it in cases of high orders.
\end{thebibliography}
\end{document}